\pdfoutput=1 % only if pdf/png/jpg images are used
\documentclass{JINST}

\usepackage{amsmath, amssymb, bm}
\usepackage{fancyhdr}
\usepackage{placeins}
\usepackage{graphicx}

\title{Scintillation Light from Cosmic-Ray Muons in\\Liquid Argon}

\author{D.~Whittington$^a$\thanks{Corresponding author.}, S.~Mufson$^b$, and B.~Howard$^a$\\
\llap{$^a$}Physics Department, Indiana University,\\
  Bloomington, IN, USA\\
\llap{$^b$}Astronomy Department, Indiana University,\\
  Bloomington, IN, USA\\
E-mail: \email{dwwhitti@indiana.edu}}
\date{\today}

\abstract{
This paper reports the results of an experiment to directly measure the time-resolved scintillation signal from the passage of cosmic-ray muons through liquid argon.  
Scintillation light from these muons is of value to studies of weakly-interacting particles in neutrino experiments and dark matter searches. 
The experiment was carried out at the TallBo dewar facility at Fermilab using prototype light guide detectors and electronics developed for the Deep Underground Neutrino Experiment.  
Two models are presented for the time structure of the scintillation light, a phenomenological model and a composite model.  Both models find $\tau_{\text{T}} = 1.52$~$\mu$s for the decay time constant of the Ar$_2^*$ triplet state.  
These models also show that the identification of the ``early'' light fraction in the phenomenological model, $F_{\text{E}}\approx$~25\% of the signal, with the total light from singlet decays is an underestimate.  
The total fraction of singlet light is $F_{\text{S}} \approx$ 36\%, where the increase over $F_{\text{E}}$ is from singlet light emitted by the wavelength shifter through processes with long decay constants.
The models were further used to compute the experimental particle identification parameter $F_{\text{prompt}}$, the fraction of light coming in a short time window after the trigger compared with the light in the total recorded waveform.  The models reproduce  quite well the typical experimental value $\sim$0.3 found by dark matter and double $\beta$-decay experiments, which suggests this parameter provides a robust metric for discriminating electrons and muons from more heavily ionizing particles.
}

\begin{document}

\section{Introduction}\label{sec:intro}

Liquid argon (LAr) is proving to be a sensitive and cost-effective detector medium for the study of weakly-interacting particles in neutrino experiments and dark matter searches. Signals generated in LAr by these particles' interactions include ionization electrons from charged daughter particles, which can be detected directly by a time projection chamber, or by photodetectors sensitive to the scintillation light from excited states in argon.  This paper reports on the properties of the scintillation light generated by cosmic-ray muons in LAr using light collectors, photodetectors, and readout electronics being developed for the Deep Underground Neutrino Experiment (DUNE).  

As charged particles pass through LAr, they can excite or ionize argon atoms.  When the excited/ionized argon atom 
pairs with a neutral argon atom, it produces in an excited argon dimer, which subsequently decays by emitting a scintillation photon.  These scintillation processes are described in Eq.~(\ref{eq:scint1}).
\begin{subequations}
\begin{align}
&\text{Ar}^* + \text{Ar} \rightarrow \text{Ar}_2^* \rightarrow 2\text{Ar} + \gamma, ~~~~~~~~~~~~~~~~~~~~~
~~~~~~~~~~~~~~~~~~~~~~~~~~~~~~~~\\ 
{\rm or}~~~~~~~~~~~~~~~~~~~~~~~~~~~~~~~~~~~~\nonumber \\
&\text{Ar}^+ + \text{Ar} \rightarrow \text{Ar}_2^+ \nonumber \\
&\text{Ar}_2^+ + e^- \rightarrow \text{Ar}_2^* \rightarrow 2\text{Ar} + \gamma. 
\end{align}
\label{eq:scint1}%
\end{subequations}
In both processes the decay of the dimer results in the emission of a vacuum ultraviolet (VUV) photon within a narrow wavelength band centered at 128~nm. 

The argon dimer $ \text{Ar}_2^*$ can be excited to either a singlet ($^1\Sigma^+_u$) or a triplet ($^3\Sigma^+_u$) state and the scintillation photons from these two states cannot be easily distinguished from one another spectroscopically. The mean lifetime of the singlet $^1\Sigma^+_u$ state is $\tau_{\text{S}}$~$\approx$~6~ns (``early'' light); the triplet $^3\Sigma^+_u$ state is significantly longer-lived, with $\tau_{\text{T}}$~$\approx$~1.5~$\mu$s (``late'' light)~\cite{bib:Hitachi1,bib:pulseShape2}. The primary objectives of this investigation are to make a precision measurement of $\tau_{\text{T}}$ for cosmic-ray muons and to characterize the relative fraction of early light to late light that they produce.
Since cosmic-ray muons are expected to behave like minimum ionizing electrons in LAr, the results reported here for muons might be more generally applied to electrons. 

Scintillation light will be useful in the analysis of experimental data from LAr detectors in many ways. For experiments with a time-projection chamber, the leading edge of the light pulse from the singlet decay provides sub-mm accuracy in the reconstruction of the absolute position of the event along the drift coordinate. For underground neutrino detectors, scintillation light can provide a trigger for proton decay, supernova neutrinos, and atmospheric neutrinos. For both neutrino and dark matter experiments, scintillation light will prove useful for particle identification. This comes about because highly ionizing particles create a higher local density of electrons than cosmic-ray muons, which induce more singlet decays.  The larger fraction of singlet to triplet decays differentiates highly ionizing particles from muons and contributes information useful to the rejection of cosmic-ray spallation backgrounds~\cite{bib:MITN2}.

Detecting the VUV scintillation photons from LAr in large neutrino detectors like DUNE is technically challenging because of the difficulty in detecting the VUV photons efficiently.  Since significant photocathode coverage of the detector is expensive, several alternative technologies have been proposed to collect the scintillation light.  One of those technologies is used in the experiment described here to collect and analyze LAr scintillation light.  With this prototype DUNE technology VUV photons are collected on the surface of a light guide.  Once these photons penetrate a short distance into the light guide, they are converted into the optical by embedded wavelength shifter.  These optical photons are then channeled to photosensors at the end of the light guide.
This technology can be made practical because LAr is a copious source of scintillation light, producing tens of thousands of VUV photons per MeV along a track~\cite{bib:ICARUS}, and pure liquid argon is transparent to its own scintillation light.  

First the experimental design is described in \S\ref{sec:ExperimentalDesign} and operating conditions are described in \S\ref{sec:operations}. Details of the various experimental subsystems are provided in \S\ref{sec:subsystems}. The system's response to scintillation light from LAr is then described in \S\ref{sec:systResponse}. Next the results of the analysis are presented in \S\ref{sec:results} and \S\ref{sec:PhysicalModel}, and their interpretation is discussed in \S\ref{sec:Discussion}. 

%~~~~~~~~~~~~~~~~~~~~~~~~~~~~~~
\FloatBarrier
\section{Experimental Design}
\label{sec:ExperimentalDesign}

The experiment took place in the 460~liter liquid argon dewar ``TallBo'' at the Proton Assembly Building (PAB) at Fermi National Accelerator Laboratory (FNAL) from November 18 through December 5, 2014.
The experimental apparatus consisted of four prototype DUNE photon detector (PD) modules immersed in LAr on  a custom paddle mount suspended from the lid.   
A PD module is made up of light guides that capture and convert VUV scintillation photons into the optical, and then channel the optical photons to silicon photomultipliers (SiPMs) at one end. 
The SiPMs were read out by prototype DUNE electronics.  
Hodoscope paddles were placed on either side of TallBo, and were used used to generate coincidence triggers that guarantee the events read out passed through the dewar and to provide basic track information. 

\begin{figure}[ht]
  \begin{center}
    \includegraphics[width=0.85\columnwidth]{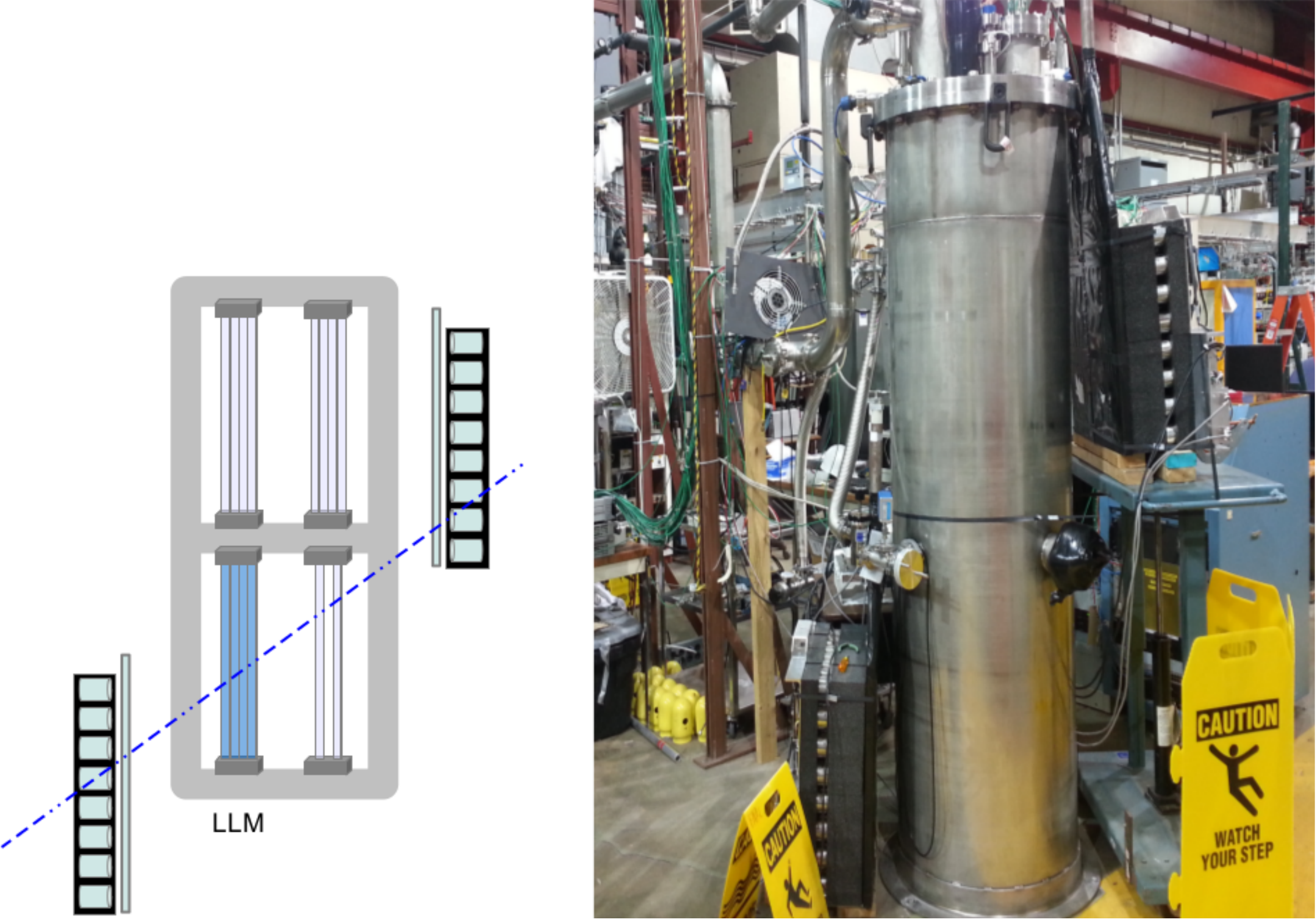}
    \caption{{\it Left:} A drawing of the experimental apparatus. Four photon detector modules in a custom frame are housed in the TallBo dewar.  Adjacent hodoscope trigger paddles flank the outside of the dewar.  A representative triggered cosmic-ray muon trajectory is superposed.  Only triggered events in the lower left PD module, LLM, are used in the analysis.  {\it Right:} A photograph of the TallBo dewar with the hodoscope modules.}
    \label{fig:Layout}
  \end{center}
\end{figure}

Fig.~\ref{fig:Layout} shows a schematic of the experiment on the left.  Four PD paddles are hung in a custom frame inside the dewar.  Two hodoscope trigger paddles, each consisting of an array of PMTs and a sheet of scintillator plastic, flank the dewar.  Data used in this analysis were acquired with the hodoscopes as shown in the left panel of Fig.~\ref{fig:Layout}.  A representative cosmic-ray muon trajectory from a hodoscope trigger is superposed.  On the right is a photograph of the TallBo dewar and the hodoscope trigger paddles.  

This experimental apparatus was primarily designed to compare the relative performance of several different prototype DUNE photon detector technologies in LAr.  For those studies the comparisons were made using both triggered events and ``free run'' or self-triggered events.  In this investigation, however, the experimental apparatus was repurposed to make a precision measurement of $\tau_{\text{T}}$ and to characterize the relative fraction of early light to late light for cosmic-ray muons.

For these studies only a subset of the triggered events were used -- those in the lower left PD module, hereafter LLM.  There are no events in the dataset from the PD paddle in the upper left in Fig.~\ref{fig:Layout} because the hodoscopes were never deployed in the high-high configuration.  Events in the upper right PD paddle and the lower right PD paddle were also excluded from the analysis.  The intent here is to determine the properties of the scintillation light from a dataset with as few systematic uncertainties as feasible.  The SiPMs in the upper right PD paddle had all been thermally cycled many times before the experiment and were quite noisy.  In the lower right there was only one light guide (3 SiPMs) which was paired with a prototype hybrid fiber array.  Both these prototype light guides were read out by a second digitizer with somewhat different operating characteristics.  Although the data from the upper and lower right PD modules were excluded from the main analysis, data collected from the 7 functional SiPMs in the upper right PD paddle and the 3 SiPMs on the light guide in the lower right PD paddle were analyzed independently.  As described below, the results found are consistent with those in the primary analysis.  

\FloatBarrier
\section{TallBo Operations}
\label{sec:operations}

To prepare for a run, the dewar was first evacuated by a turbo pump to help reduce contamination from residual gasses and then back-filled with gaseous argon.  The gaseous argon was next replaced with ultra-high-purity (UHP) LAr that passed through a molecular sieve and copper filter on the fill line that had been regenerated just prior to the run.  The contaminants that most affect LAr scintillation light are O$_2$, N$_2$, and H$_2$O, which can both quench scintillation light and decrease the argon transparency at 128~nm~\cite{bib:MITN2}.  UHP LAr is typically delivered with low levels of these contaminants, and the regenerated filtering apparatus is very effective at further removing O$_2$ contamination.  In addition, studies at the Materials Test Stand (MTS) in PAB showed that the light guides do not outgas measurable O$_2$ and only 2-3~ppb of H$_2$O when submerged in LAr.   The MTS does not test for outgassing by N$_2$.  Tests at the MTS have also shown that the Teflon-jacketed cables do not outgas contaminants.  

The O$_2$, N$_2$, and H$_2$O concentrations were monitored during the run.  They could not all be monitored continuously, however, because the O$_2$ and N$_2$ monitors cannot sample from the LAr volume simultaneously.  Although the O$_2$ and H$_2$O monitors can operate simultaneously, in practice the H$_2$O monitor was often needed for the adjacent MTS.  Since O$_2$ does not outgas significantly from the components, it was only sampled at the beginning, middle, and end of the run.  The H$_2$O was sampled at the middle and end of the run.  The N$_2$ was monitored when the O$_2$ and H$_2$O were not being monitored.  

The O$_2$ contamination was found to be $\sim$40~ppb when sampled for 18--24~hr at the beginning and middle of the run.  At the end of the run, after the O$_2$ monitor had been continuously operating for 3 days, the O$_2$ contamination leveled off at $\sim$40~ppb.  At concentrations $<$100~ppb, the effects of O$_2$ contamination on scintillation light in LAr are negligible~\cite{bib:O2Contamination}.  The N$_2$ contamination from day~2 -- day~7 ranged from $\sim$80--190~ppb.  From day~8 -- day~14 the N$_2$ contamination leveled off at $\sim$80~ppb.  At these concentrations, the N$_2$ does not affect the scintillation light~\cite{bib:MITN2,bib:N2Contamination}.  On 11/25 the H$_2$O contamination was measured to be $\sim$8~ppb.  From day~15 -- day~18 the H$_2$O concentration leveled off at $\sim$5~ppb.  The effects of H$_2$O contamination on LAr scintillation light are not well studied.  Measurements in gaseous argon~\cite{bib:H2O}, however, show that the neither the fast nor the slow component of scintillation light is significantly affected by H$_2$O contamination below 10~ppb.  It is presumed here that the H$_2$O concentration in this experiment does not affect our results.

Once filling was complete, the vessel was sealed and subsequently maintained at a positive internal pressure of 10 psig to ensure no contamination entered the dewar from the outside during the run.  A liquid-argon condenser on TallBo reliquified gas from the ullage and returned it to the dewar in order to maintain a constant liquid level inside.  Since only cables and connectors were in the ullage, no contamination should be introduced when reliquifying the LAr.

\section{Experiment Subsytems}
\label{sec:subsystems}

\subsection{Hodoscope Paddles and Trigger}
\label{sec:Trigger}

Two hodoscope modules were installed on opposite sides of the TallBo dewar to select single-track cosmic-ray muons passing through the LAr volume. These hodoscope modules were loaned from the CREST balloon-based cosmic-ray experiment~\cite{bib:CREST}.  Each  hodoscope module consists of 64 2-inch diameter barium-fluoride crystals, coated with TPB and arranged in an 8$\times$8 array. Each crystal is monitored by a single PMT. 

Since the hodoscope modules were designed to detect bremsstrahlung photons from high-energy electrons bending in the Earth's magnetic field, they are very sensitive to extraneous photon activity around our experiment. To reject these $\gamma$ showers, a pair of plastic scintillator panels covering the entire face of a hodoscope module were placed between each hodoscope module and the TallBo dewar. These panels were each individually read out by a PMT. The SiPM readout was triggered by four-fold coincidence logic that required at least one hit in both hodoscope modules as well as one hit in their adjacent scintillator planes. This trigger guarantees that each event contains at least one charged particle passing through the liquid argon. Events were further filtered offline to reject showers by requiring one and only one hit in each hodoscope module. Single-track events crossing from one side of the frame to the other were rejected in order to exclude any tracks that could pass through a light guide.

\FloatBarrier
\subsection{Light Guides}
\label{sec:LightGuides}

\begin{figure}[ht]
  \begin{center}
    \includegraphics[width=0.85\columnwidth]{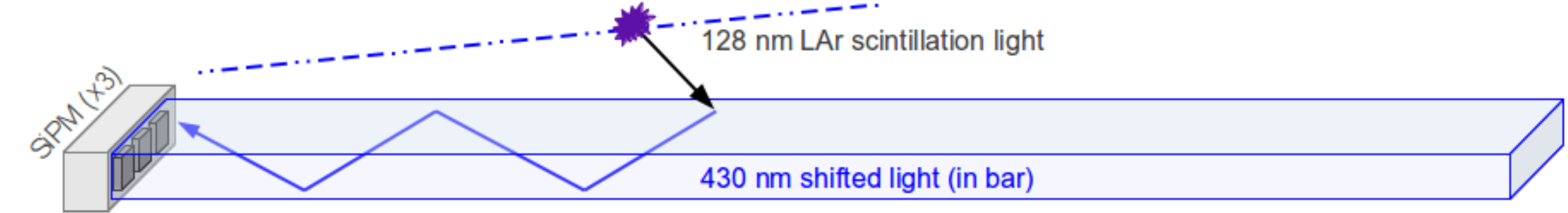}
    \caption{Schematic of a light guide with 3 SiPMs on one end. VUV scintillation photons impinging on the light guide surface are converted by the WLS into visible photons inside the bar, which then propagate down the light guide via total internal reflection. The photons that reach the readout end are detected by the SiPM array with high efficiency.}
    \label{fig:lightguide}
  \end{center}
\end{figure}

The experiment was carried out with the four light guides in the lower left of Fig.~\ref{fig:Layout}.
A schematic drawing of a light guide with its photosensors is shown in Fig.~\ref{fig:lightguide}. The light guides are manufactured from cast acrylic or polystyrene bars of dimensions 50.8 cm $\times$ 2.54 cm $\times$ 0.6 cm that have wavelength shifter (WLS) embedded in them. The concept is described in Ref.~\cite{bib:MITbars}. 
The wavelength shifter converts VUV scintillation photons to $\sim$430~nm photons inside the bar, with a reasonable efficiency for converting a VUV to an optical photon~\cite{bib:gehman}.  A fraction of the waveshifted optical photons are internally reflected to the light guide's end where they are detected by SiPMs whose QE is well matched to the 430~nm waveshifted photons. The light guides were made with one of two wavelength shifters: the conventional TPB (1,1,4,4-tetraphenyl-1,3-butadiene) and the less expensive alternative bis-MSB (1,4-bis-(o-methyl-styryl)-benzene). Preliminary studies with a VUV monochromator show that the two wavelength shifters compare favorably in their waveshifting efficiency~\cite{bib:baptistaJINST}.  Similar results are found here.

The light guide designs used in this experiment are given in Table~\ref{tab:lightGuides}.  Two different processes were used for incorporating the wavelength shifter into the light guides -- ``dip-coating'' and ``doping.''  In addition, two different wavelength shifters -- TPB and bis-MSB -- were used in manufacturing the light guides.  Below is a brief description of the two processes used to manufacture the light guides.
\begin{table}[ht]
  \begin{center}
    \caption{Light guide designs in LLM.}
    \vspace{0.2em}
    \label{tab:lightGuides}
    \begin{tabular}{| c  l  l  |}
      \hline
      \hline
      Light Guide &Technology   & WLS      \\
      \hline
      A & acrylic, dip-coated   & TPB      \\
      B & acrylic, dip-coated   & bis-MSB  \\
      C & acrylic, doped        & bis-MSB  \\
      D & polystyrene, doped    & TPB      \\
      \hline
      \hline
    \end{tabular}
  \end{center}
\end{table}

\medskip

\noindent ``Dip-Coating'': Light guides A and B were produced using a dip-coating technology developed at MIT~\cite{bib:MITbars} as an alternative to hand-painting, 
although the process used here differs in several ways.
The light guides were made from blanks of commercially available Lucite-UTRAN cast UVT acrylic sheet that was laser-cut into bars of the proper size.  Lucite-UTRAN has the longest attenuation length of the UVT acrylics tested~\cite{bib:mufsonJINST}.  Before the wavelength shifter was applied, the acrylic bars were annealed at 80$^\circ$C for one hour.  
To embed the WLS in the light guide, either TPB or bis-MSB was first dissolved in the organic solvent dichloromethane (CH$_2$Cl$_2$).  For light guides A and B there were 2 gm of wavelength shifter dissolved in 1,000 gm of dichloromethane (DCM).  The bar was dipped into the WLS mixture for 15 seconds and then removed.  The bar was then hung in the dark for at least two hours to dry.  Once dry, the ends of the bars were flycut. 

\medskip

\noindent ``Doping'': Light guides C and D were cut from a sheet of acrylic or polystyrene cast with TPB or bis-MSB mixed into the plastic. The doped acrylic sheets were manufactured by Astra Products\footnote{http://astraproducts.com}. The polystyrene sheets were manufactured by Eljen Technology\footnote{http://www.eljentechnology.com}. The sheets had either 1\% TPB or 1\% bis-MSB by mass added during their respective proprietary casting processes, which distributes WLS throughout the volume. If more than 1\% of either wavelength shifter is added, WLS crystallizes out and degrades the light guide's performance.  Since VUV photons have a very short penetration depth in acrylic, this manufacturing method uses far more WLS than necessary. (The penetration depth of 128~nm photons in polystyrene is $<$~100~nm, as estimated from Ref.~\cite{bib:Buck}, and it is expected that polystyrene and acrylic have similar absorption in the VUV.) On the other hand, light guides function more efficiently when their surfaces are flat and the casting process results in very flat surfaces, which is a mitigating factor if the prime consideration is efficiency for the detection of VUV photons. 

\FloatBarrier
\subsection{Photodetectors and Photodetector Readout}
\label{sec:PhotoSensorsReadout}

At the end of each light guide are 3 SensL MicroFB-60035-SMT SiPMs\footnote{\label{sensl}sensl.com}. Each SiPM has an active area of $6 \times 6$ mm$^2$.  They are made up of an array of 18,960 microcell photodiodes, each of which is 35 $\mu$m on a side, and the microcell filling factor on the chip is 64\%. The SiPMs are reverse-biased at 24.5 V.

The operating characteristics of these SiPMs have been determined by the manufacturer\footnotemark[3] down to 230~K.  For operation in LAr, the SiPMs need to be characterized at 87~K.  Since these measurements require SiPM dark spectra, the SiPMs were read out while immersed in liquid nitrogen (LN2). The dark measurements were made in LN2 rather than LAr because LN2 does not scintillate, so no systematics are introduced by scintillation light from cosmic-rays or radioactive impurities. The 10~K difference in temperature between LAr and LN2 is not expected to impact the results.

The signals from the 12 SiPMs in the LLM were processed by a 12-channel SiPM Signal Processor (SSP) module that was designed and built by the HEP Electronics Group at Argonne National Laboratory (ANL) for the DUNE photon detection system.  Each of the readout channels consists of a differential voltage amplifier and a 14-bit, 150 MSPS ADC that digitizes the signals with negligible dead time. The ADC has a full-scale dynamic range of 2~V, corresponding to approximately 1000 photoelectrons (pe's) at a typical SiPM gain of $3.5\!\times\!10^6$. The amplifier input impedance is 100~$\Omega$ with an overall digitizer gain of 1850 V/A. Each ADC count is equivalent to 2V/2$^{14} \times (18.5~V/V)^{-1} = 6.60\!\times\!10^{-3}~$mV. The SSP was designed to resolve single pe pulses and achieves a resolution of 18\% FWHM.

An FPGA in the SSP implements an independent data processor for each channel. The processing includes a leading edge discriminator to detect events, amplitude analysis algorithms for measuring the peak and the integral of the waveform, a pulse pileup detection algorithm, and a constant fraction discriminator.  The bin width of each sample is 6.67~ns.  Internal studies have shown that this readout system is capable of better than 3~ns timing resolution.  

\begin{figure}[ht]
  \begin{center}
    \includegraphics[width=0.9\columnwidth]{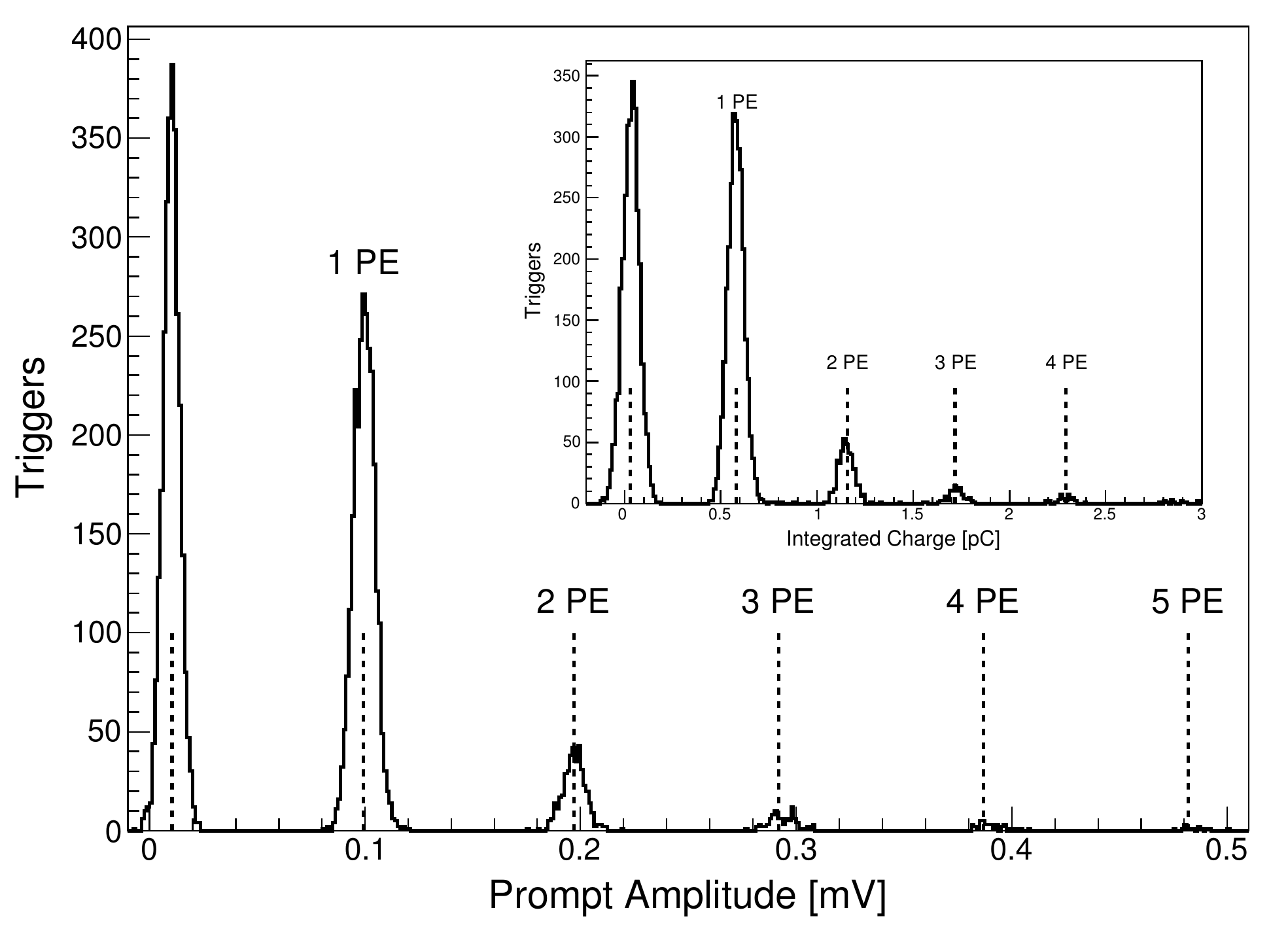}
    \caption{Dark spectra for the prompt signal and the integrated signal ({\it inset}) for the SensL MicroFB-60035-SMT SiPM A-1 in LN2.  The prompt amplitude is the average signal in the first 133~ns of the waveform. The integrated signal has been calibrated to charge in pC. The excellent single-pe resolution and the linearity of the response are apparent.}
    \label{fig:NoiseFigs}
  \end{center}
\end{figure}

Fig.~\ref{fig:NoiseFigs} shows dark spectra acquired in 300~s with the SSP for one instance of a SensL MicroFB-60035-SMT SiPM in LN2, the SiPM on light guide A in position 1, SiPM A-1. The SiPM was biased at $V_{\text{b}}$ = 24.5 V.
The prompt amplitude histogram shows the average signal in the first 133~ns (20 bins) of the waveforms for the dark events.  The inset histogram gives the integrated charge in the waveforms over their full range.  Both histograms show the excellent single-microcell resolution and the linearity of the response for this SiPM. The integrated charge in each waveform is determined by first summing the waveform amplitude (in ADC counts) over the $N$ $\times$ 6.67~ns time bins in the waveform, where typically $N = 580$. The integrated charge collected in the waveform is calculated by converting this sum ($s$) to charge ($c$), by $c~\text{[pC]} = s \times (6.60\!\times\!10^{-6}~\text{[V/ADC count]}) / 100\,\Omega \times (6.67\!\times\!10^{-9}~\text{s})/10^{-12}$.

The gains for the SiPMs were determined by dividing the mean integrated charge in the 1~pe peak in their dark spectra by the charge of an electron.  
The gains for the SiPMs at $V_{\text{b}}$ = 24.5~V in the LLM are given in Table~\ref{tab:SiPM-Noise}.  
\begin{table}[ht]
  \begin{center}
    \caption{Dark noise characteristics of SiPMs in LLM in LN2 at $V_{\text{b}}$ = 24.5 V.}
    \vspace{0.2em}
    \label{tab:SiPM-Noise}
    \begin{tabular}{| l |  c  c  c  c  c  |}
      \hline
      \hline
      SiPM  & Gain & Noise & Cross-Talk  & Rise Time & Recovery Time \\
            &      & [Hz]  & Probability &   [ns]    &       [ns]    \\
      \hline
      A-0 &  3.5$\times10^6$ & 11 & 0.17 & 13 & 484 \\
      A-1 &  3.6$\times10^6$ &  9 & 0.21 & 12 & 483 \\
      A-2 &  3.6$\times10^6$ &  9 & 0.18 & 13 & 494 \\
      \hline
      B-0 &  3.5$\times10^6$ &  9 & 0.22 & 11 & 460 \\
      B-1 &  3.6$\times10^6$ &  8 & 0.19 & 14 & 491 \\
      B-2 &  3.5$\times10^6$ & 10 & 0.19 & 14 & 502 \\
      \hline
      C-0 &  3.4$\times10^6$ &  8 & 0.22 & 13 & 492 \\
      C-1 &  3.6$\times10^6$ & 10 & 0.20 & 12 & 473 \\
      C-2 &  3.5$\times10^6$ &  8 & 0.18 & 11 & 449 \\
      \hline
      D-0 &  3.6$\times10^6$ &  8 & 0.19 & 13 & 468 \\
      D-1 &  3.5$\times10^6$ &  9 & 0.22 & 13 & 479 \\
      D-2 &  3.5$\times10^6$ &  8 & 0.22 & 13 & 491 \\
      \hline
      \hline
      mean    & 3.5$\times10^6$ & 9 & 0.20 & 13 & 481 \\
      st.dev. & 6.5$\times10^4$ & 1 & 0.02 &  1 &  16 \\
       \hline
      \hline
   \end{tabular}
  \end{center}
\end{table}
The dark noise rates for the SiPMs in the LLM were determined from their dark spectra by summing the number of triggers with prompt amplitudes $\geq$~0.5~pe and dividing the sum by the 300~s data acquisition time.  For SiPM A-1 in Fig.~\ref{fig:NoiseFigs}, where 0.5~pe $\approx$~0.05~mV, the noise rate is 11~Hz.  The dark rates for the remaining SiPMs are given in Table~\ref{tab:SiPM-Noise}.  At LN2 temperatures, these SiPMs clearly have very low noise rates.  As the bias voltage increases to $V_{\text{b}}$ = 28.5 -- 29~V, the noise rate increases to $\sim$100~Hz.  Above $V_{\text{b}} \approx$~29~V, the noise rate begins to rise exponentially, reaching $\sim$1~kHz at $V_{\text{b}}$ = 31~V.  For this experiment, the SiPMs were biased at 24.5~V, where the gain is high but before the noise rate has begun its rapid rise. 

The cross talk probability given in Table~\ref{tab:SiPM-Noise} is computed as the ratio of triggers with integrated charge $\geq$~1.5~pe to triggers with integrated charge $\geq$~0.5~pe.  Cross talk events occur when a photon 
emitted during the electron avalanche in one pixel is re-absorbed by another pixel elsewhere on the SiPM and induces a second avalanche in immediate coincidence with the first.  The definition of cross talk used here also includes events which occur on a time scales of less than a few tens of nanoseconds when electrons migrate from one avalanche into a neighboring pixel and induce after-pulsing, a second avalanche in delayed coincidence with the first.  Studies of similar devices from a different manufacturer\cite{crossTalk} suggests the 5~$\mu$s dark noise waveforms in Fig.~\ref{fig:NoiseWaveform_A_1} include almost all after-pulsing events.  The cross talk (and after-pulsing on short time scales) probability at 77~K is $\sim$2--3 times greater than its value at room temperature\footnotemark[3].

The SiPM response to incident photons is a rapid rise to a maximum value followed by an exponential relaxation back to the baseline. The rise and recovery times of the SiPMs in Table~\ref{tab:SiPM-Noise} were measured using the average response to dark noise events in LN2 (described in detail in \S~\ref{subsec:single_pe}). For each SiPM, an error function was fit to the rising edge and an exponential function is fit to the tail ($t$~>~300~ns). The rise time was defined as the difference in time between the points where the error function takes on values of 10\% and 90\% of its maximum. The time constant of the exponential function describes the SiPM recovery time.  Although the SiPMs have fast rise times like PMTs, they exhibit a much longer recovery time.  

\FloatBarrier 
\section{Scintillation Light from Cosmic-Ray Muons}
\label{sec:systResponse}

\FloatBarrier
\subsection{Cosmic-Ray Muon Waveforms and the Illumination Function}
\label{subsec:cosmicWaveforms}

\begin{figure}[ht]  
\begin{center}
    \includegraphics[width=0.80\columnwidth]{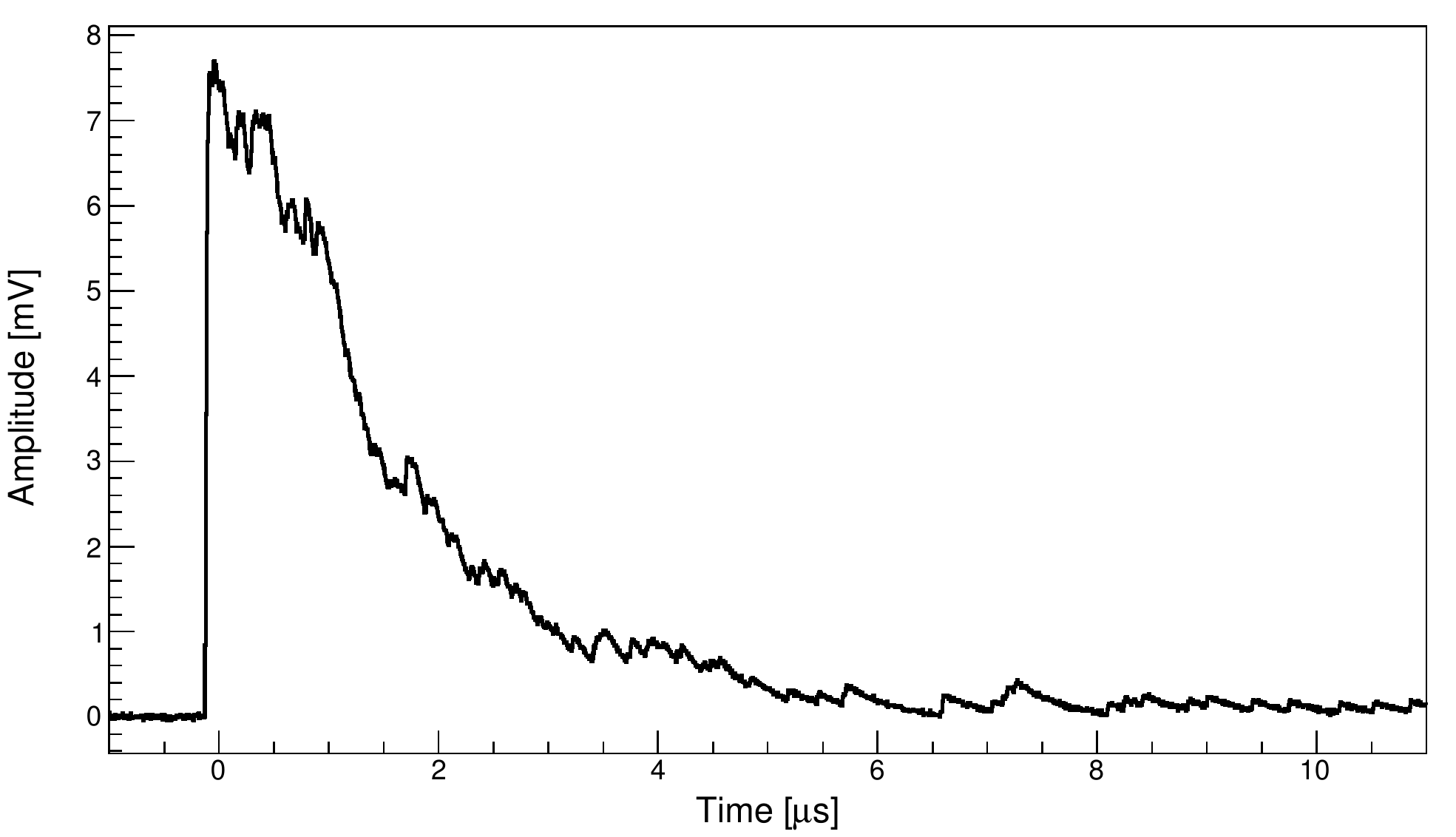}
    \caption{An example waveform generated by the scintillation light from a single-track cosmic-ray muon in the LLM of TallBo.  This hodoscope-selected track was recorded by SiPM A-1. The waveform is the superposition of the multi-photon pulse from early light and the subsequent few or single-photon pulses from the late light convolved with the detector's single-pe response function.  The  leading edge of this waveform is the sum of approximately 20 single-pe pulses.}
    \label{fig:WaveEx}
  \end{center}
\end{figure}

Fig.~\ref{fig:WaveEx} shows an example of a waveform from a single-track cosmic-ray muon in the LLM of TallBo as viewed by SiPM A-1.  
It is the convolution of the time sequence of scintillation photons incident on the light guide, or the ``illumination function'', with the response function of the detector.  Qualitatively, the illumination function for this waveform is made up of an early, multi-photon pulse arriving within the first few ns, which is then followed by a number of single or few photon pulses that stretch over more than 10 $\mu$s.  Since an SiPM is made up of 18,960 microcells, the probability of near-simultaneous photon hits on a single microcell is quite small.  Therefore, this waveform can be regarded as the superposition of a series of single photon hits on random microcells, each of which is convolved with the detector's single photoelectron response function.  This approximation is adopted in the analysis that follows.  

Like the waveform in Fig.~\ref{fig:WaveEx}, all the waveforms studied here consist of 1950 $\times$ 6.67~ns samples, with the first 300 pre-trigger samples (2~$\mu$s) used to determine the baseline.  The start time of the event, $t_0 = 0$, is the arrival time of the sharp leading edge of the early-light photons.  

%\FloatBarrier
\subsection{Determination of the Mean Illumination Function}
\label{subsec:meanIlluminationFunction}

The objective of this investigation is to find the illumination functions characterizing the scintillation light from cosmic-ray muons in LAr for the different light guides studied here and then to build models that best describe the behavior of these functions.  Let $w(t_i)$ be the amplitude of the pedestal-corrected waveform in time bin $t_i$ from a single track cosmic-ray muon event, as in Fig.~\ref{fig:WaveEx}.  This waveform represents the convolution of the illumination function, $i(t_i)$, with the single photoelectron response function, $f_{pe}(t_i)$, 
\begin{equation}
\label{eq:conv}
 w(t_i)  = f_{pe} \ast i.
\end{equation}
In this equation, $f_{pe}(t_i)$ includes the effects of cross talk and after-pulsing associated with each photoelectron.

The exact waveform seen by an SiPM varies from event to event and depends on both the path length and orientation of the muon track in the dewar.  Our assumption is that the detector response to scintillation of the liquid argon can be deduced from the mean pedestal-subtracted waveforms for the ensemble of $N$ waveforms seen by the the SiPMs.  For SiPM $k$, the mean pedestal-subtracted waveform, $ \langle w^k(t_i)\rangle$, is given by 
\begin{equation}
\langle w^k(t_i)\rangle  = \sum_{j=1}^N w^k_j(t_i)/N,
\label{eq:meanWaveform}
\end{equation}
Since each SiPM $k$ has its own single photoelectron response function, the analysis has been carried out separately for each SiPM.  Further, cross talk and after-pulsing are stochastic processes, so the analysis was done with the mean single photoelectron response function, $F^k_{pe}(t_i) \equiv \langle f_{pe}(t_i) \rangle$. The mean illumination function on SiPM $k$, $I^k(t_i) \equiv \langle i^k(t_i) \rangle$ is found by deconvolving 
\begin{equation}
\langle w^k(t_i)\rangle  = F^k_{pe} \ast I^k.
\label{eq:deconvolution}
\end{equation}

\subsubsection{Mean Single Photoelectron Response Function}
\label{subsec:single_pe}

\begin{figure}[ht]
  \begin{center}
    \includegraphics[width=0.85\columnwidth]{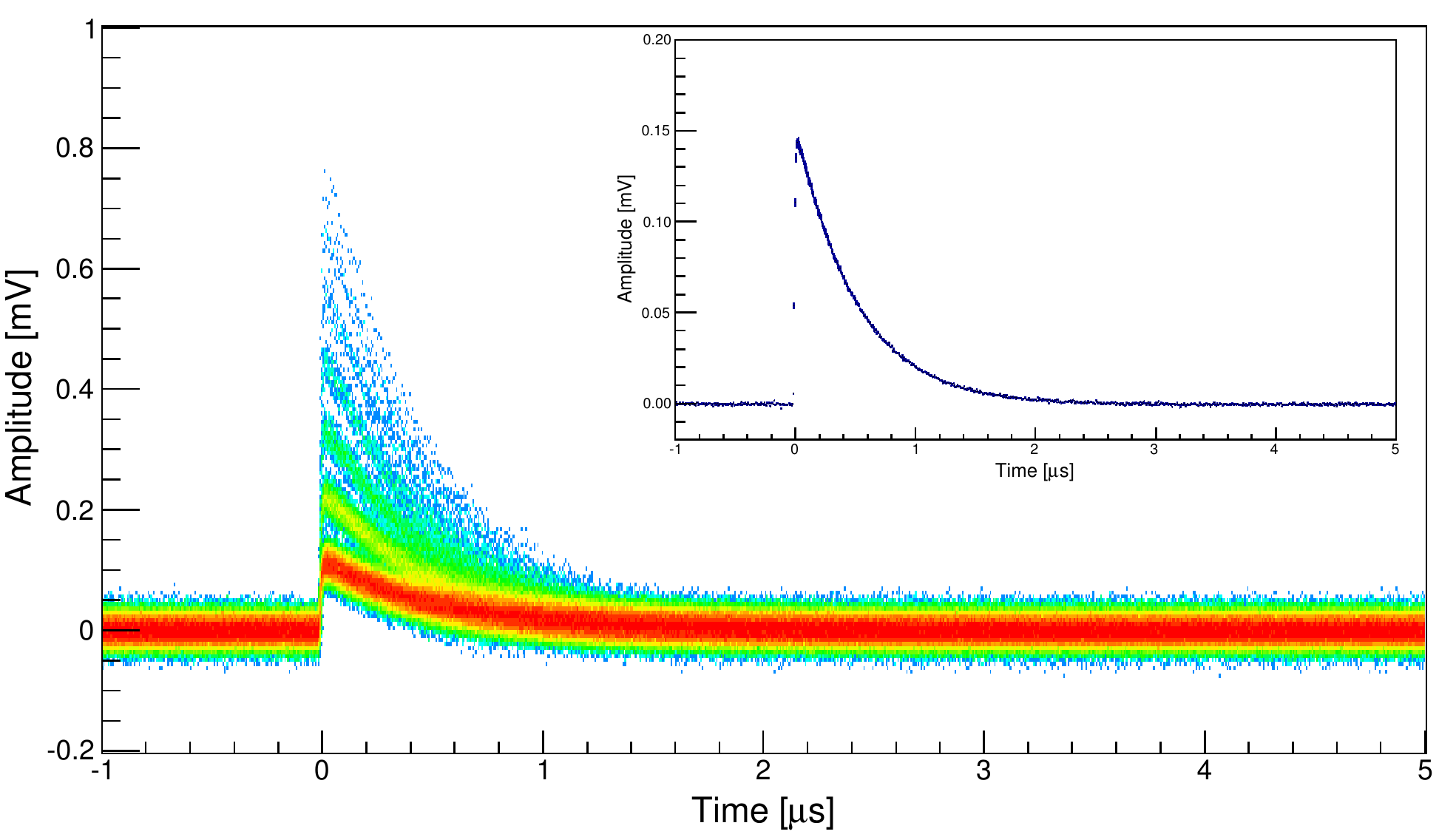}
  \caption{All dark noise triggers for SiPM A-1 in 300~s of data acquisition.  {\it inset}:  The mean single pe response function, $F^{\text{A-1}}_{pe}(t_i) $ for all dark noise triggers in SiPM A-1.}  
    \label{fig:NoiseWaveform_A_1}
  \end{center}
\end{figure}

Fig.~\ref{fig:NoiseWaveform_A_1} shows the superposition of the waveforms for all LN2  dark noise triggers recorded by SiPM A-1 in 300~s of data acquisition.  Each digitized dark noise waveform consists of 900 samples, which includes 150 
pre-trigger samples.  The total waveform is 6~$\mu$s long. The inset figure shows the mean single photoelectron 
response function, $F^{\text{A-1}}_{pe}(t_i)$, used in Eq.~(\ref{eq:deconvolution}) to find the mean illumination function for SiPM A-1.  It was computed from this set of waveforms by averaging the contents in each time bin using using the ROOT\footnote{http://root.cern.ch} class TProfile.  Mean single photoelectron response functions have been computed in a similar way for the remaining 11 SiPMs in the LLM.  These mean single photoelectron response functions represent the mean response of the SiPM to a single photon hit.  

\subsubsection{Mean Cosmic-Ray Muon Waveform}
\label{subsec:mean_cosmic}

\begin{figure}[ht]
  \begin{center}
    \includegraphics[width=0.85\columnwidth]{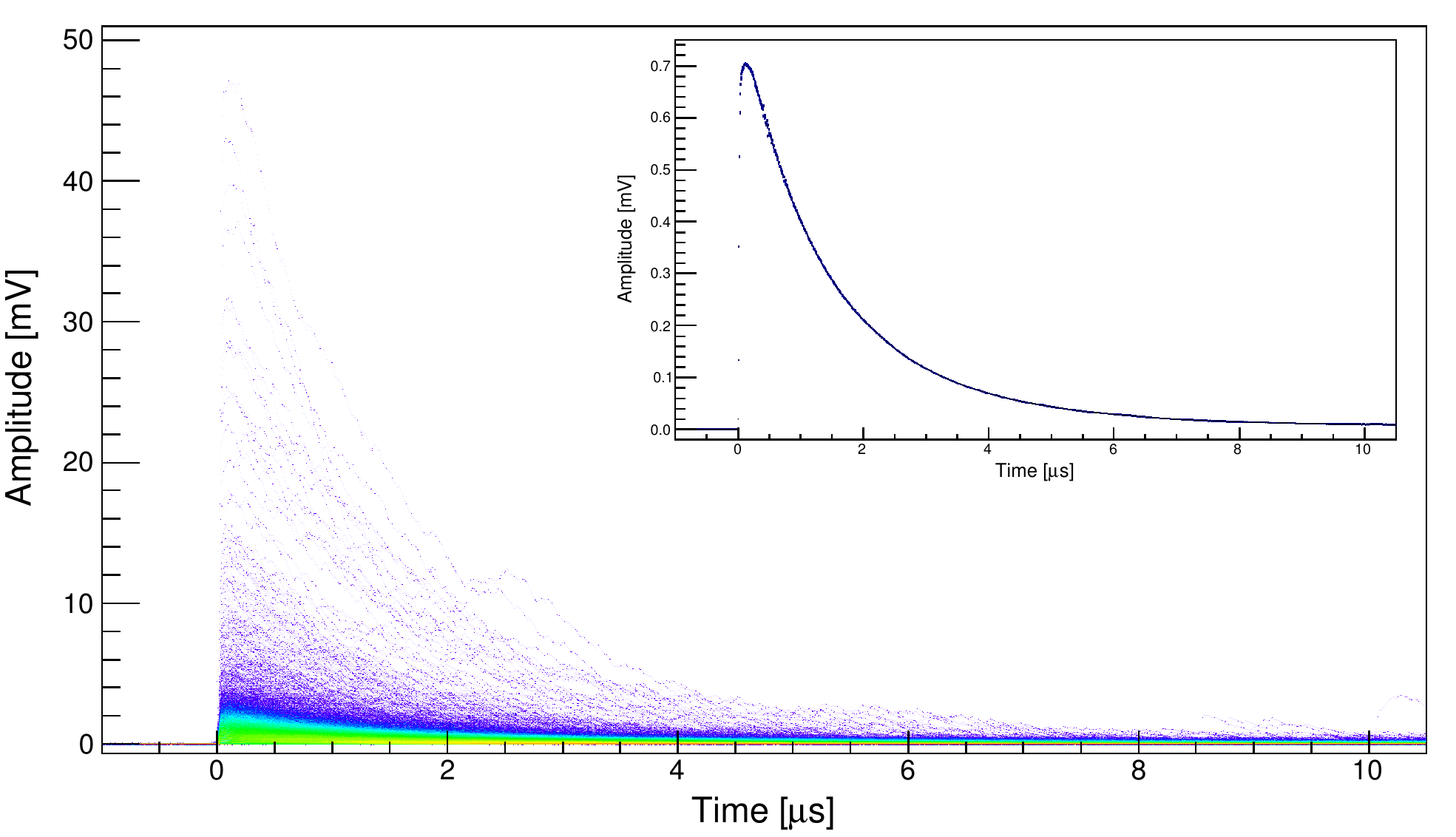}
    \caption{The superposition of all cosmic-ray muon waveforms $w(t_i)$ in the LLM of TallBo.  These hodoscope-triggered waveforms were recorded by SiPM A-1.  The mean waveform $\langle w^{\text{A-1}}(t_i)\rangle$ derived from the cosmic-ray triggers is shown in the inset.  The response of liquid argon to cosmic-ray muons, or the mean illumination function $I^k(t_i)$, is obtained by deconvolving the mean single photoelectron response function from the mean cosmic-ray muon waveform.}
    \label{fig:AllWaveformsEx}
  \end{center}
\end{figure}

Fig.~\ref{fig:AllWaveformsEx} shows the superposition of all waveforms from single-track cosmic-ray muons recorded by SiPM~A-1 in the LLM of TallBo that pass the four-fold coincidence and offline single-track selection cuts.  The inset shows the mean cosmic-ray muon waveform, $\langle w^{\text{A-1}}(t_i)\rangle$, which was computed by averaging the contents in each time bin for all the waveforms using using the ROOT class TProfile.
The mean single photoelectron response function, $F^{\text{A-1}}_{pe}(t_i)$, is deconvolved from the mean waveform $\langle w^{\text{A-1}}(t_i)\rangle$ to find the mean illumination function $I^{\text{A-1}}(t_i)$.  

The mean cosmic-ray waveforms for the remainder of the SiPMs in the LLM were computed and analyzed in a similar way.  More than 35,000 single cosmic-ray muon tracks were used to construct the mean waveform for each of the SiPMs.    

\subsubsection{Deconvolution Procedure}
\label{subsec:deconvolution}

The deconvolution of the mean single photoelectron response function from the mean cosmic-ray muon waveform uses the data for the mean response function $F^{\text{A-1}}_{pe}(t_i) $, {\it e.g.} (Fig.~\ref{fig:NoiseWaveform_A_1},inset), with the Gold deconvolution algorithm implemented in the ROOT TSpectrum class to obtain the 12 time-dependent profiles of incident scintillation photons $I^k(t_i)$.  Each mean illumination function $I^k(t_i)$ represents the time-resolved structure of the signal from a cosmic-ray muon incident on the SiPM $k$. 

\begin{figure}[ht]
  \begin{center}
    \includegraphics[width=0.85\columnwidth]{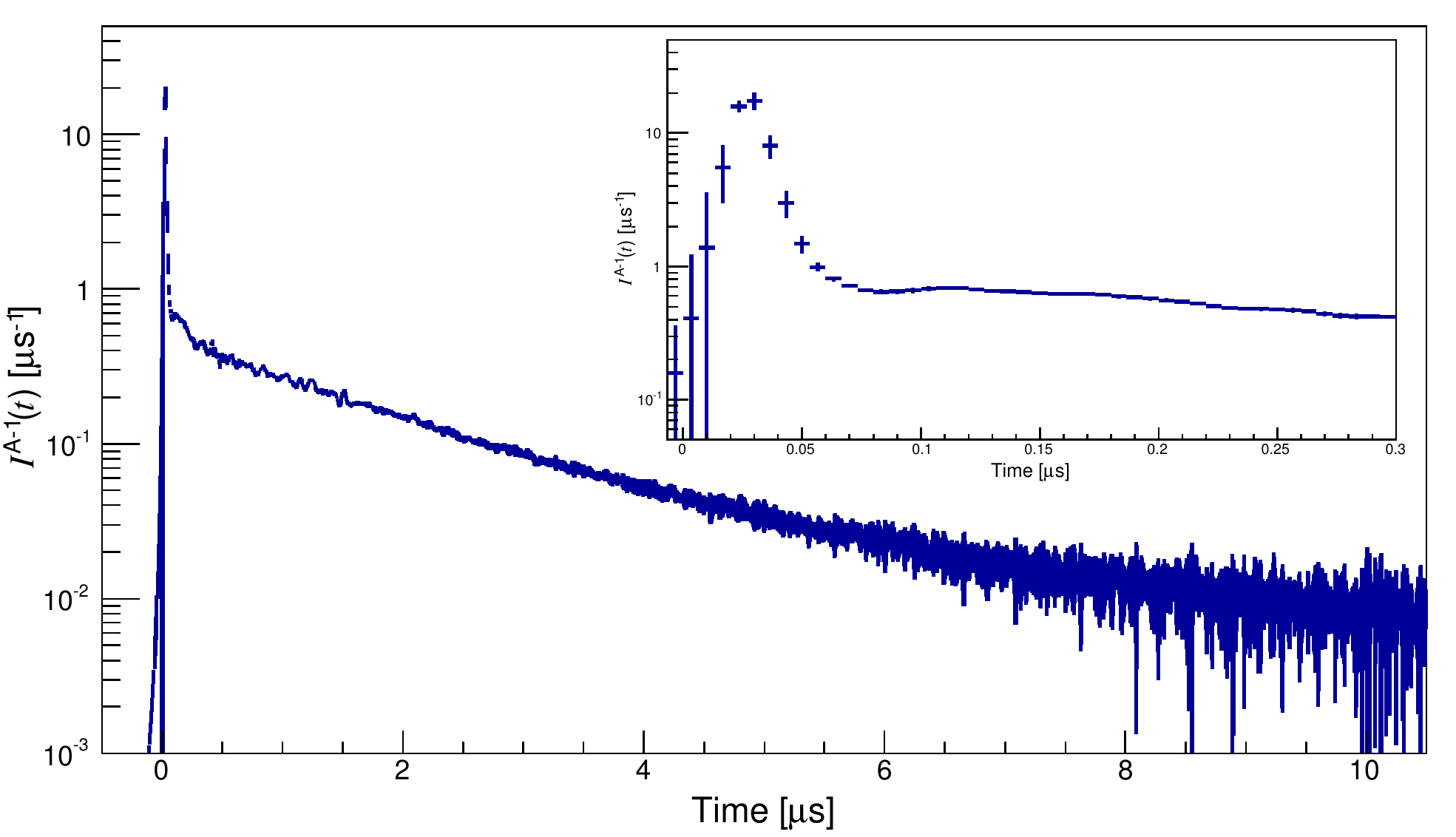}
    \caption{The deconvolved time-dependent profile of incident scintillation photons $I^k(t_i)$ for all single track waveforms recorded by SiPM $k = $~A-1. $I^k(t_i)$ is shown for the full waveform and for the first 300~ns in the inset.}
    \label{fig:DecoEx}
  \end{center}
\end{figure}

Fig.~\ref{fig:DecoEx} shows the deconvolved $I^{\text{A-1}}(t_i)$ for all single track waveforms recorded by SiPM~A-1. 
Statistical uncertainties were obtained from ten pseudo-experiments in which the bin contents of $\langle w^k(t_i) \rangle$ and $F^k(t_i)$ were varied randomly according to their standard deviations.  Each of these pseudo-experiments was deconvolved in the same manner as the mean cosmic-ray waveforms and the resulting standard error on $I^k(t_i)$ was assigned as the statistical uncertainty.

%\FloatBarrier
\section{Results}
\label{sec:results}

A phenomenological model has been developed to describe the behavior of the illumination functions that is based on a sum of exponentially modified Gaussian (EMG) functions.  The time dependence of the scintillation signal should be well approximated by a sum of exponential probability distributions with various amplitudes and lifetimes. To incorporate the smearing effects due to detector resolution and the few-ns WLS response, a Gaussian function is convolved with the exponential response.  The EMG function, $ \mathcal{E}$, is the convolution of these Gaussian and exponential probability density functions~\cite{bib:gale} and is given by
\begin{equation}
  \mathcal{E}(t;\tau,w,t_m) = 
  \frac{1}{\tau} \times \exp{\left[ \frac{1}{2}\left(\frac{w}{\tau}\right)^2 - \left(\frac{t-t_m}{\tau}\right)\right]} 
  \times \frac{1}{2} \left[1 + {\rm erf} \left( \frac{z}{\sqrt{2} } \right)\right],
  \label{eq:ExpModGauss}
\end{equation}
%\begin{equation}
%  \mathcal{E}(t;\tau,w,t_m) = 
%  \frac{1}{\tau} \times \frac{1}{2} \left[1 + {\rm erf} \left( \frac{z}{\sqrt{2} } \right)\right]
% \times \exp{\left[ \frac{1}{2}\left(\frac{w}{\tau}\right)^2 - \left(\frac{t-t_m}{\tau}\right)\right]},
%  \label{eq:ExpModGauss}
%\end{equation}
\noindent where
\begin{equation}
  z = \left(\frac{t-t_m}{w}-\frac{w}{\tau}\right). \nonumber
\end{equation}
In this function, $w$ is the width of the Gaussian, $t_m$ is the Gaussian mean, and $\tau$ is the parameter characterizing the exponential falloff from the peak. The error function component of this normalized function accounts for the rapid rise in the illumination function.  The functional dependence on $\exp(-t/\tau)$ accounts for the exponential fall-off.

A single EMG function can adequately characterize the rapid rise in the illumination function.  It cannot, however, provide a satisfactory fit to the long tails in the illumination functions, an example of which is seen Fig.~\ref{fig:DecoEx} where the structure in the tail is apparent.  
For SiPM $k$, the $n$-component model for its illumination function is given by
\begin{equation}
I^k(t_i) = \sum_{j=1}^n A_j\; \mathcal{E}_j(t_i;\tau_j^k,w^k,t_m^k), 
\label{eq:response}
\end{equation}
where the $A_j$ are normalization constants, the parameters $w$ and $t_m$ that describe the functional rise are constrained to common values for all the components, and the parameters $\tau_j^k$that model the exponential fall-off is free to take different values for each of the components.  
There are $2n + 2$ fit parameters in this model for each SiPM $k$.

The fits were performed using MINUIT in ROOT.  They were carried out over a range of 10 $\mu$s for light guides A, B, and C.  For light guide D, where the statistics were more limited in the final microsecond, the fits were carried out over a range of 8.5 $\mu$s.

\subsection{Multi-Component Models for the Illumination Functions}
\label{sec:4comp}

The de-excitation of the $\text{Ar}_2^*$ dimer has two decay components, a fast ``early light'' signal from singlet state decays and a slower ``late light'' signal from triplet state decays.  The simplest model for the illumination functions $I^k(t_i)$ is therefore one with two components.  The fits for two-component models, however, all resulted in values of $\chi^2/N_{\text{DF}}$  between 4.8 -- 6.7 for the SiPMs.  These $\chi^2/N_{\text{DF}}$ values show that two-component models are not adequate to describe the behavior of the illumination functions.  

Next, three and four-component models for the illuminaton functions were investigated.  
For the SiPMs on the acrylic light guides A, B, and C, four-component models were necessary for $\chi^2/N_{\text{DF}} < 1$.  These models are parameterized by an early light (``E'') component, an intermediate light component (``I'') component, a late light (``L'') component, and a fourth (``4'') component with a long time constant ($>6~\mu$s).  These models have 10 fit parameters -- four amplitudes $A$, four decay constants $\tau$, and the two Gaussian parameters $w$ and $t_m$.  For the SiPMs on the polystyrene light guide, the fourth component has an amplitude consistent with zero and three component models gave acceptable fits.  These models have 8 fit parameters.
The detailed results of these fits for each SiPM are given in the tables in Appendix~A.
\begin{figure}[ht]
  \begin{center}
    \includegraphics[width=.49\columnwidth]{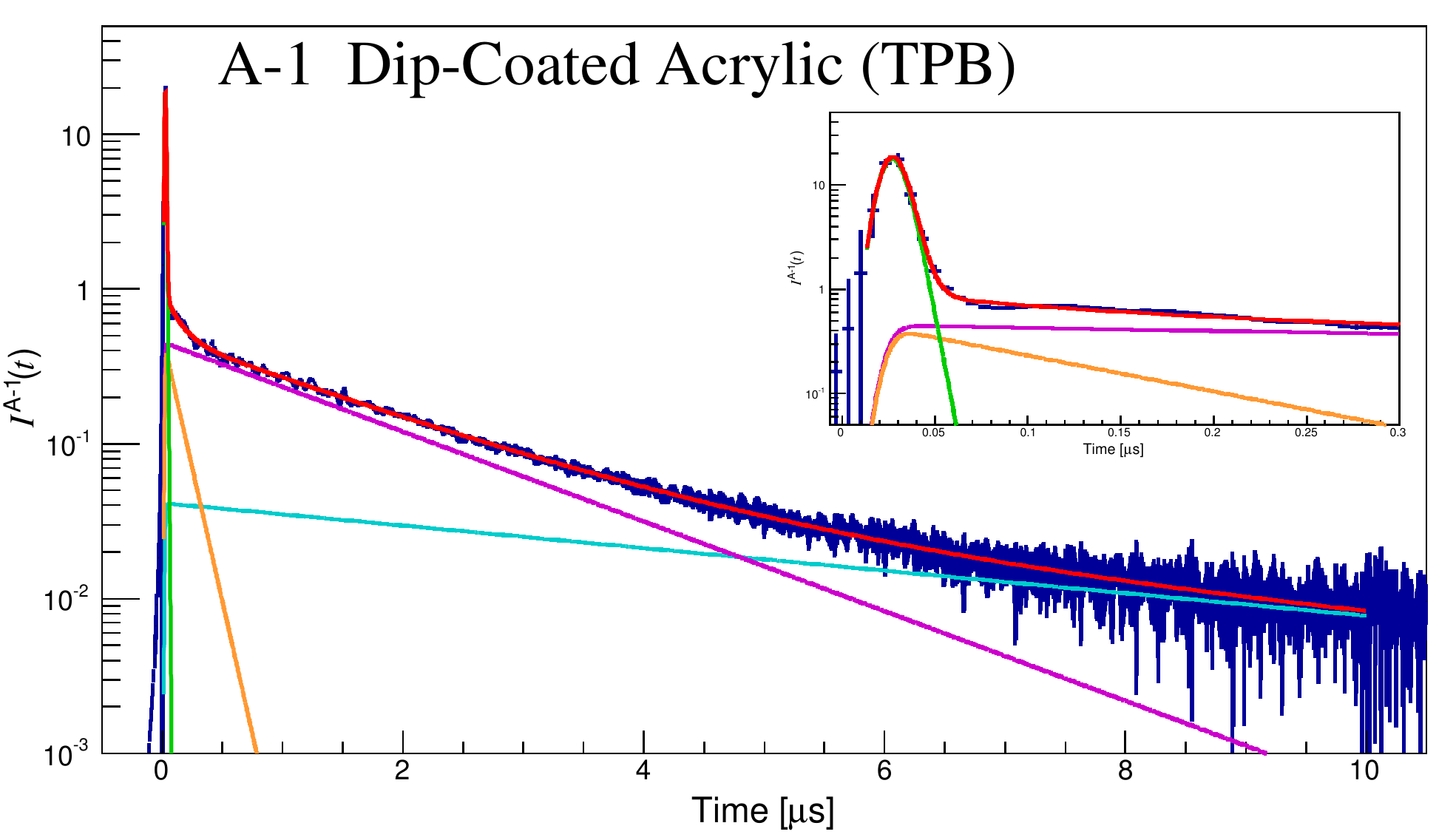} \hspace{0.1em}
    \includegraphics[width=.49\columnwidth]{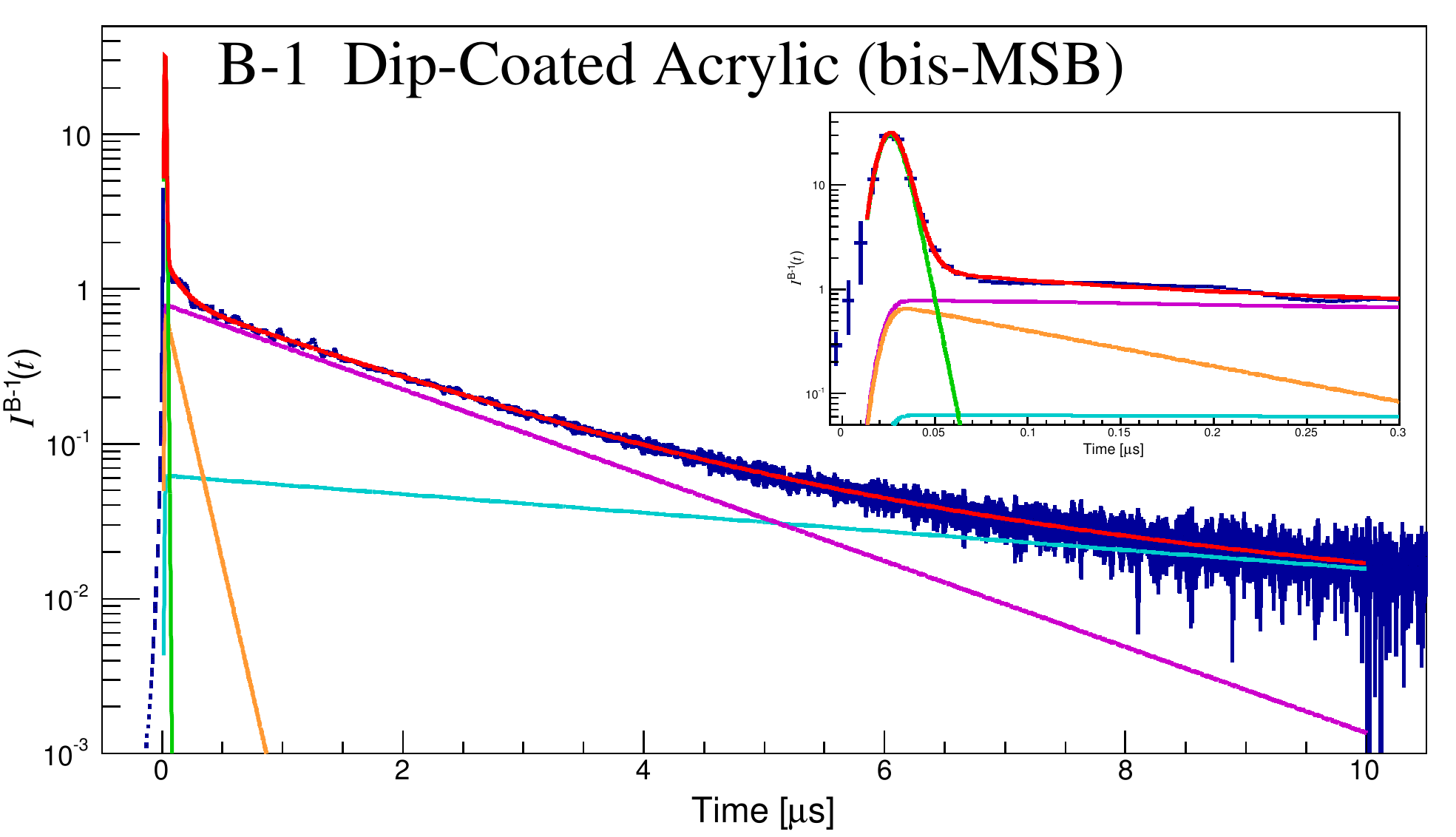} \\
    \vspace{1em}
    \includegraphics[width=.49\columnwidth]{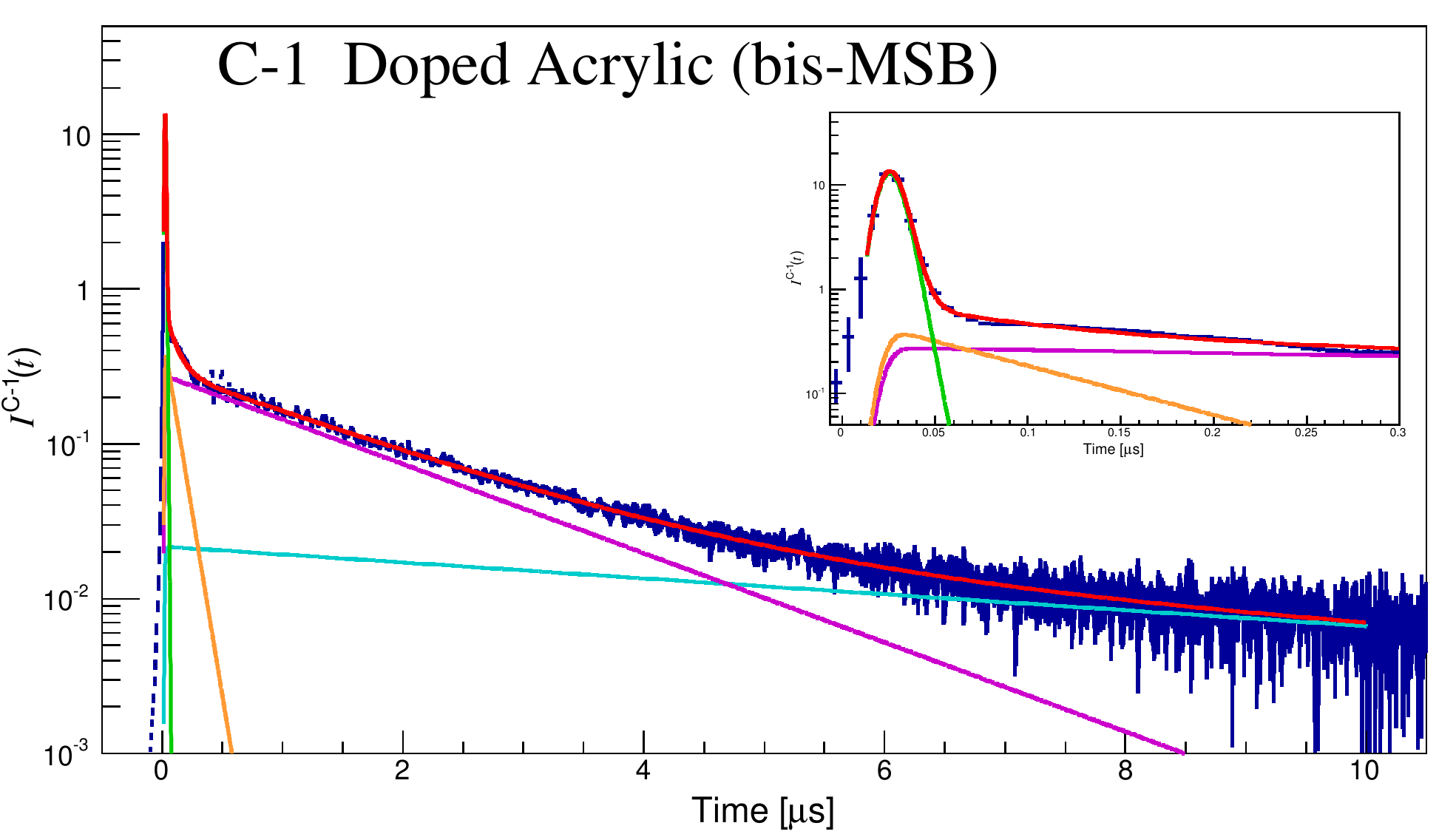} \hspace{0.1em}
    \includegraphics[width=.49\columnwidth]{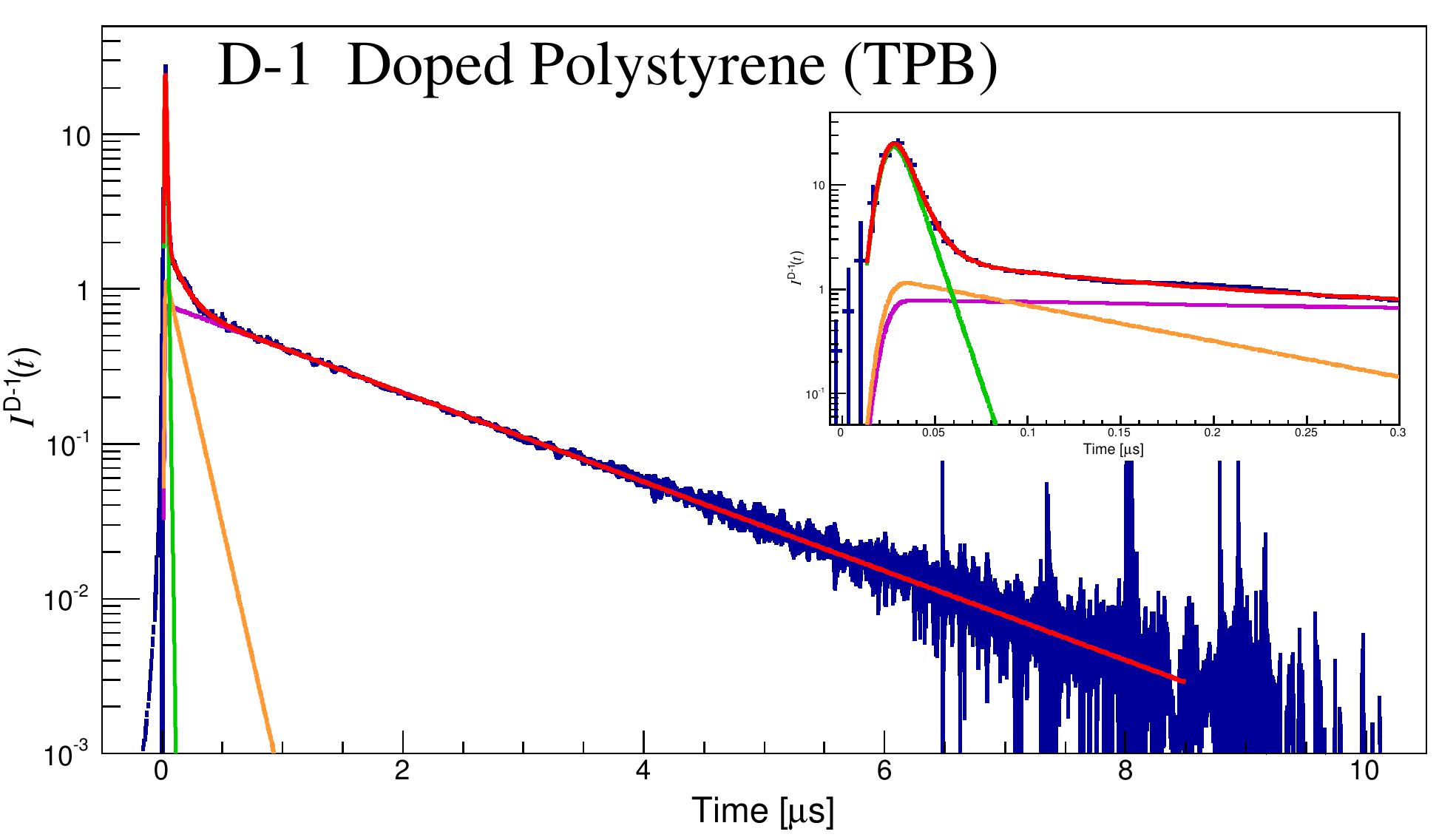}
    \caption{The illumination function $I^k$ observed by SiPM 1 for light guides A, B, C, and D.  The multi-component fits are superposed in red.  In addition, each component is shown separately: green for the early ``E'' component, orange for the intermediate ``I'' component, magenta for the late, ``L'' component, and cyan for the fourth ``4'' component.  For SiPM D-1, the amplitude of the fourth component is consistent with zero.}
    \label{fig:4comp-ScintFits}
  \end{center}
\end{figure}
The statistical errors on the fit parameters are assigned by MINUIT.   

Fig.~\ref{fig:4comp-ScintFits} graphically shows the fits for SiPM~1 on each of the four light guides superposed on their respective illumination functions based on the model parameters given in Appendix~A.  
This figure shows that in addition to the expected contributions from the early and the late light,  all models require a contribution from an ``intermediate'' component that accounts for the illumination function in the range 50 -- 300~ns.  Evidence for a signal from an intermediate component has been previously reported~\cite{bib:Hitachi1,bib:pulseShape2,bib:N2Contamination,bib:pulseShape}.  The SiPMs on the acrylic light guides A, B, and C also require the fourth component which is absent in the SiPMs on the polystyrene light guide D.  These results suggest that the fourth component is related to the composition of the light guide and is not associated with after-pulsing on microsecond time scales.  
After-pulsing is a property of the photodetectors and would be expected on all light guides.  In fact, there is no evidence in Fig.~\ref{fig:4comp-ScintFits} for after-pulsing on any time scale from tens of nanoseconds to 10 microseconds.  

Table~\ref{tab:4comp} gives the means of fit parameters for the three SiPMs on each light guide and their standard deviations from the tables in Appendix~\ref{sec:AppendixA}.  .
\begin{table}[ht]
  \begin{center}
    \caption{Means of the fit parameters and their standard deviations for the 3 SiPMs on each light guide.}
    \vspace{0.2em}
    \label{tab:4comp}
    \begin{tabular}{| c r@{\,$\pm$\,}l r@{\,$\pm$\,}l r@{\,$\pm$\,}l r@{\,$\pm$\,}l r@{\,$\pm$\,}l |}
    \hline
    \hline
    Light Guide & \multicolumn{2}{c}{$\tau_{\text{E}}$ [ns]} & \multicolumn{2}{c}{$\tau_{\text{I}}$ [ns]} & \multicolumn{2}{c}{$\tau_{\text{L}}$ [ns]} & \multicolumn{2}{c}{$\tau_4$ [ns]} & \multicolumn{2}{c|}{$F_{\text{E}}$ [\%]} \\
    \hline
    A & 4.4 & 1.9 &  127 & 10 & 1517 &  38 & 6830 & 1230 & 25.0 & 2.7 \\
    B & 4.8 & 1.0 &  127 & 11 & 1528 &  39 & 6350 & 910 & 23.6 & 2.7 \\
    C & 4.7 & 1.3 & 104 & 9 & 1527 &  44 & 8120 & 1620 & 27.1 & 3.1 \\
    D & 8.0 & 1.0 &  130 & 6 & 1524 & 7 & \multicolumn{2}{c}{---} & 25.8 & 2.1 \\
    \hline
    \hline
    \end{tabular}
  \end{center}
\end{table}
Listed in Table~\ref{tab:4comp} are the lifetimes for the early, intermediate, late, and fourth components of the illumination functions, along with the parameter $F_{\text{E}}$, a measure of the relative intensity of the early scintillation light component compared with the total signal.  Each of these parameters is discussed in turn below:

\medskip

\noindent Early component lifetime ($\tau_{\text{E}}$): The early component lifetimes found for light guides A, B, and C are consistent with one another and with values previously reported~\cite{bib:Hitachi1,bib:pulseShape2,bib:N2Contamination,bib:pulseShape,bib:LArScint} for the lifetime of the singlet $\text{Ar}_2^*$ state.  Light guide D has a statistically significant longer singlet lifetime.  The fact that light guide D is made from polystyrene and A, B, and C are made from acrylic probably plays a role in this difference.  Polystyrene is a known scintillator~\cite{bib:berlman} and detectors made with polystyrene doped with WLS (e.g.\ PPO, PTP, etc.) are common tools in particle physics experiments. Such doped polystyrene scintillators have been shown to emit visible light with a two-component structure including components with characteristic lifetimes in the range 6--27~ns~\cite{bib:PolystyScint}. If the decay constant associated with the combination of polystyrene and WLS in light guide D is somewhat longer than $\tau_{\text{E}}$ found for light guides A, B, and C, as suggested in \cite{bib:Swank}, this substructure could not be distinguished with the time resolution of this experiment and a longer decay constant for the early component would be measured.  

\medskip

\noindent Intermediate component lifetime ($\tau_{\text{I}}$): The decay constants for the intermediate component found here are longer than those reported in~\cite{bib:Hitachi1,bib:N2Contamination}.  Features with similar time scales to the intermediate component have been reported in~\cite{bib:pulseShape2,bib:pulseShape}. Although it is not known whether this component is the result of a physics process or is instrumental in origin, the variation in reported values and discrepancy between light guide C and light guides A, B, and D here suggest an instrumental explanation.

\medskip

\noindent Late component lifetime ($\tau_{\text{L}}$): The fit results are consistent with one another within a narrow range. The lifetime of this component indicates that it is due to the decay of triplet $\text{Ar}_2^*$ states in the LAr. This result is the first direct measurement of the triplet state lifetime $\tau_{\text{L}}$ determined using cosmic-ray muons. $\tau_{\text{L}}$ has been reported for electrons, $\gamma$'s, $\alpha$'s and fission fragments, where measurements range from 1.2--1.6~$\mu$s~\cite{bib:Hitachi1, bib:pulseShape2, bib:O2Contamination,bib:N2Contamination,bib:pulseShape,bib:LArScint}.
The values for $\tau_{\text{L}}$ in Table~\ref{tab:4comp} fall in this range.  The $e^-$ measurements in the literature for the most part do not have overlapping error bars, which makes it challenging to assess whether $\tau_{\text{L}}$ for muons and electrons are consistent with one another, as expected.  

\medskip

\noindent Fourth component lifetime  ($\tau_4$):  This component has not been previously reported.  It is seen in light guides A, B, and C, but not in light guide D.  Its presence is clear in the deviation from an exponential fall-off at $t > 6$~$\mu$s in the deconvolved illumination functions $I^k$ for light guides A, B, and C in Fig.~\ref{fig:4comp-ScintFits}). Since light guides A, B, and C are made from acrylic, while light guide D is made from polystyrene, this component also appears to be instrumental in origin.

\medskip

\noindent Early light fraction ($F_{\text{E}}$): $F_{\text{E}} = A_{\text{E}}/(A_{\text{E}}+A_{\text{I}}+A_{\text{L}}+A_4)$ for the light guides. The consistency of $F_{\text{E}}$ in the four-component models suggests that $F_{\text{E}}$ is a property of the scintillation light from LAr and is independent of light guide technology. Since the fraction of $\text{Ar}_2^*$ dimers excited to the singlet state depends on the ionization properties of the incident particles, $F_{\text{E}}$ has been used as a parameter distinguishing muons from more heavily ionizing particles in LAr~\cite{bib:Hitachi1, bib:pulseShape2,bib:pulseShape}.  This fraction is the dominant contribution to the prompt signal experimenally observed.

\subsection{Fits to All SiPMs}
\label{sec:AllSiPMs}

There were an additional 10 functional SiPMs that acquired track information during the experiment but were excluded from the analysis because they were noisy or their data were acquired with different electronics.  Since this experiment was expected to reach its best precision with quiet SiPMs and the same readout electronics, the results reported in Table~\ref{tab:4comp} were obtained by using only the high quality, restricted data set from the 12 SiPMs in Table~\ref{tab:SiPM-Noise}.  However, the data from the excluded 10 SiPMs were also analyzed using the same methods as the primary data set as a check for consistency with the results in Table~\ref{tab:4comp}.  Seven of these additional SiPMs were mounted on acrylic light guides and three were on polystyrene.

\begin{figure}[ht]
  \begin{center}
    \includegraphics[width=0.43\columnwidth]{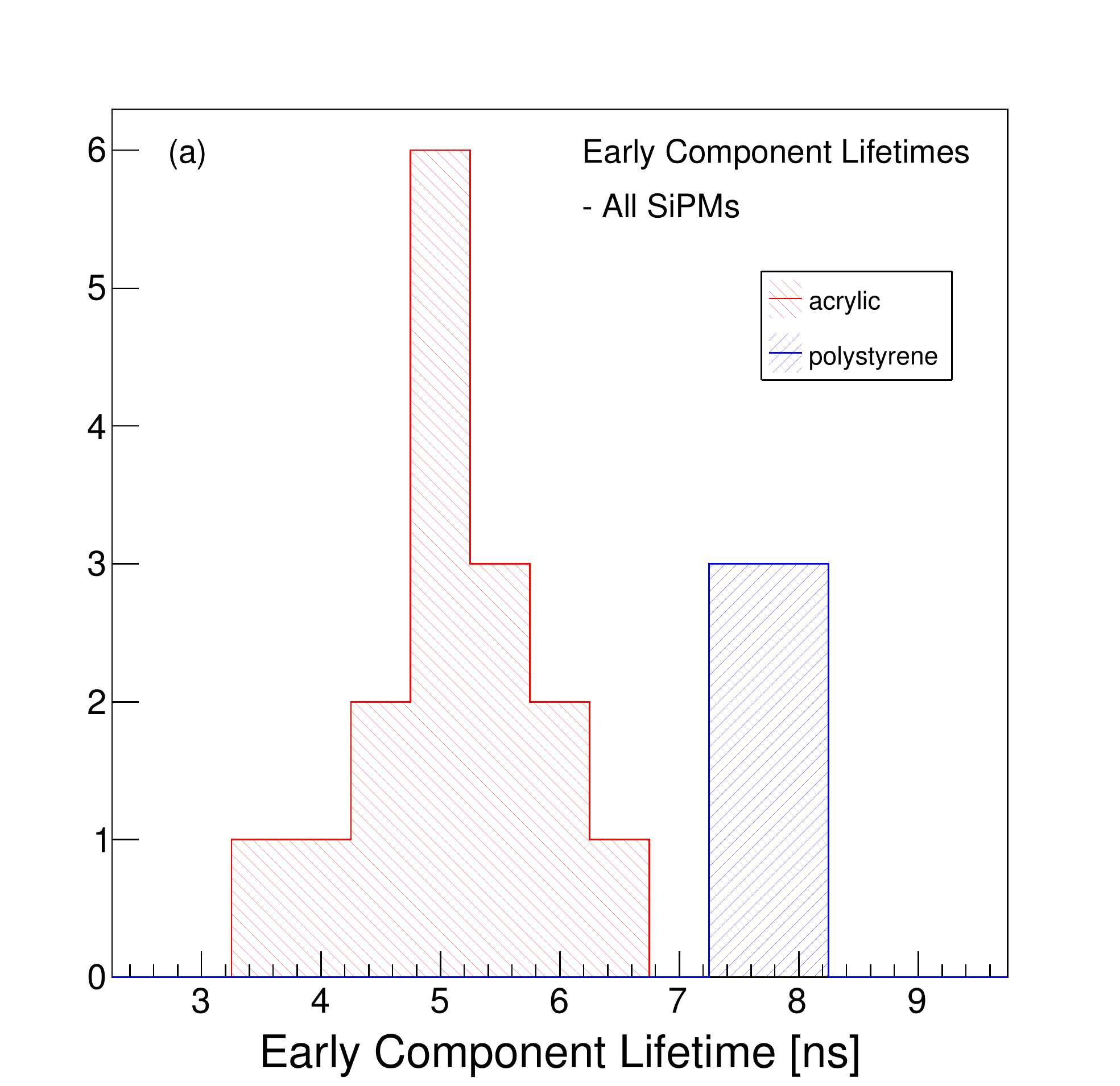} \hspace{0.02\columnwidth}
    \includegraphics[width=0.43\columnwidth]{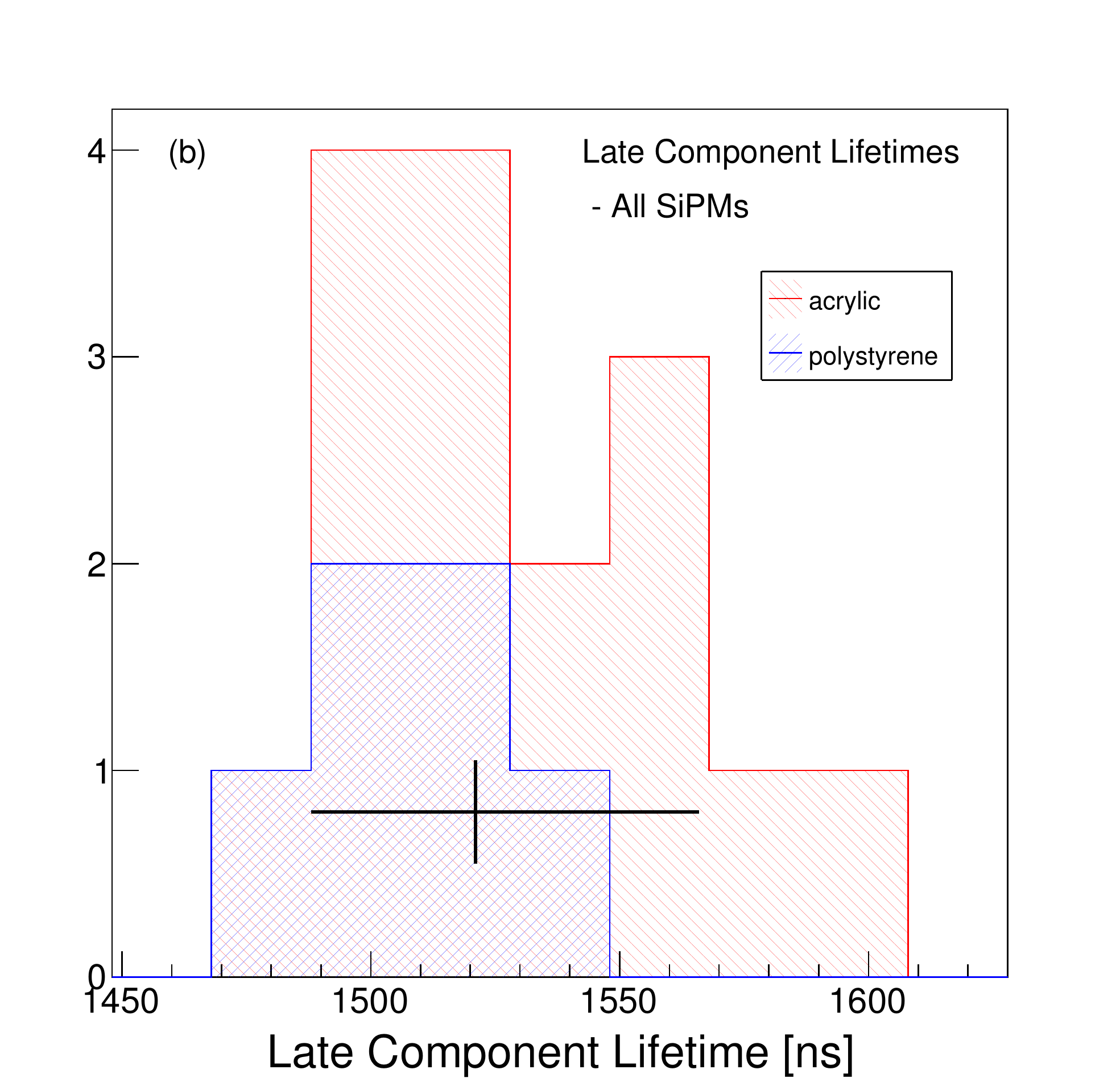} \\
    \includegraphics[width=0.43\columnwidth]{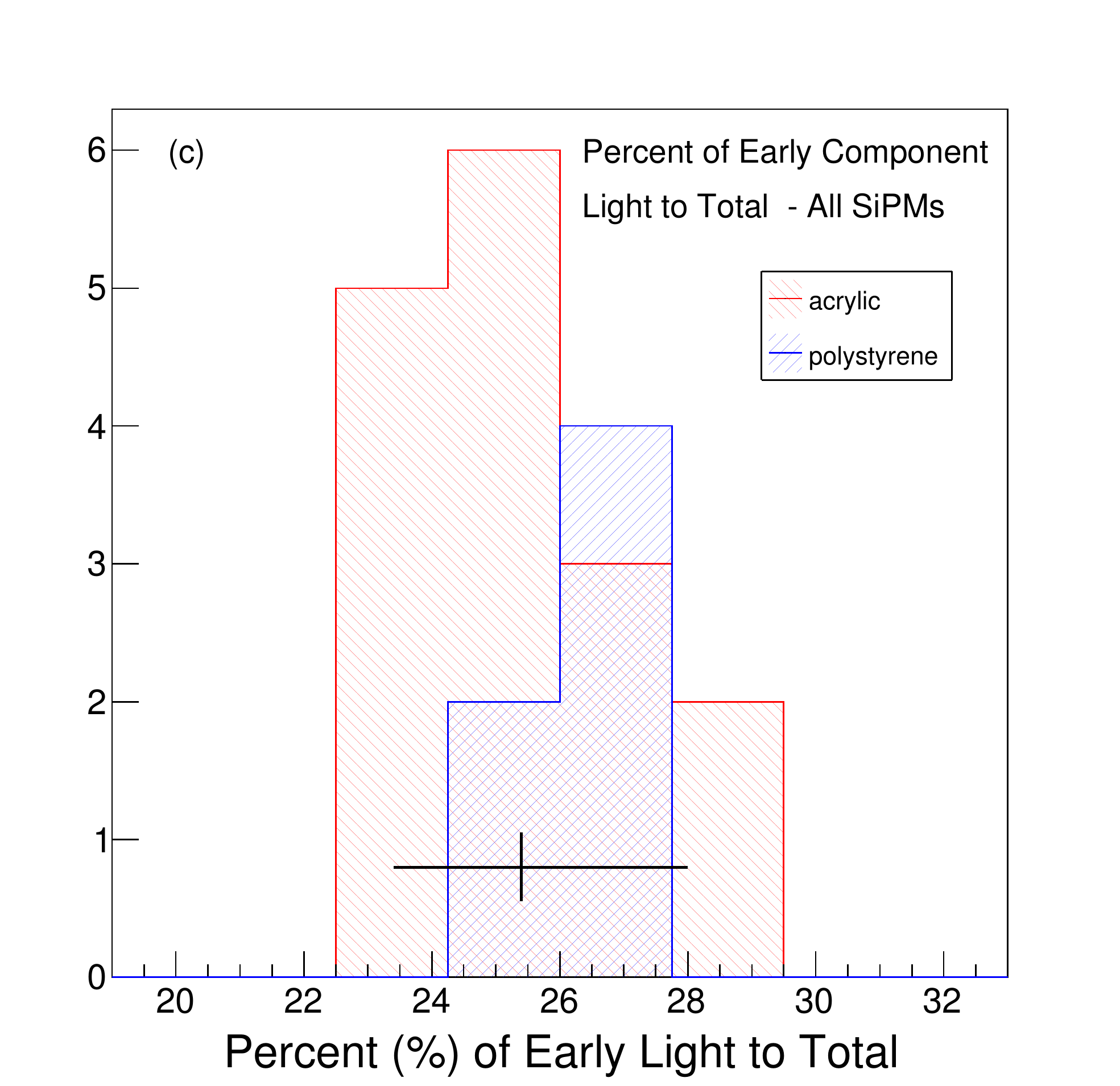} \hspace{0.02\columnwidth}
    \includegraphics[width=0.43\columnwidth]{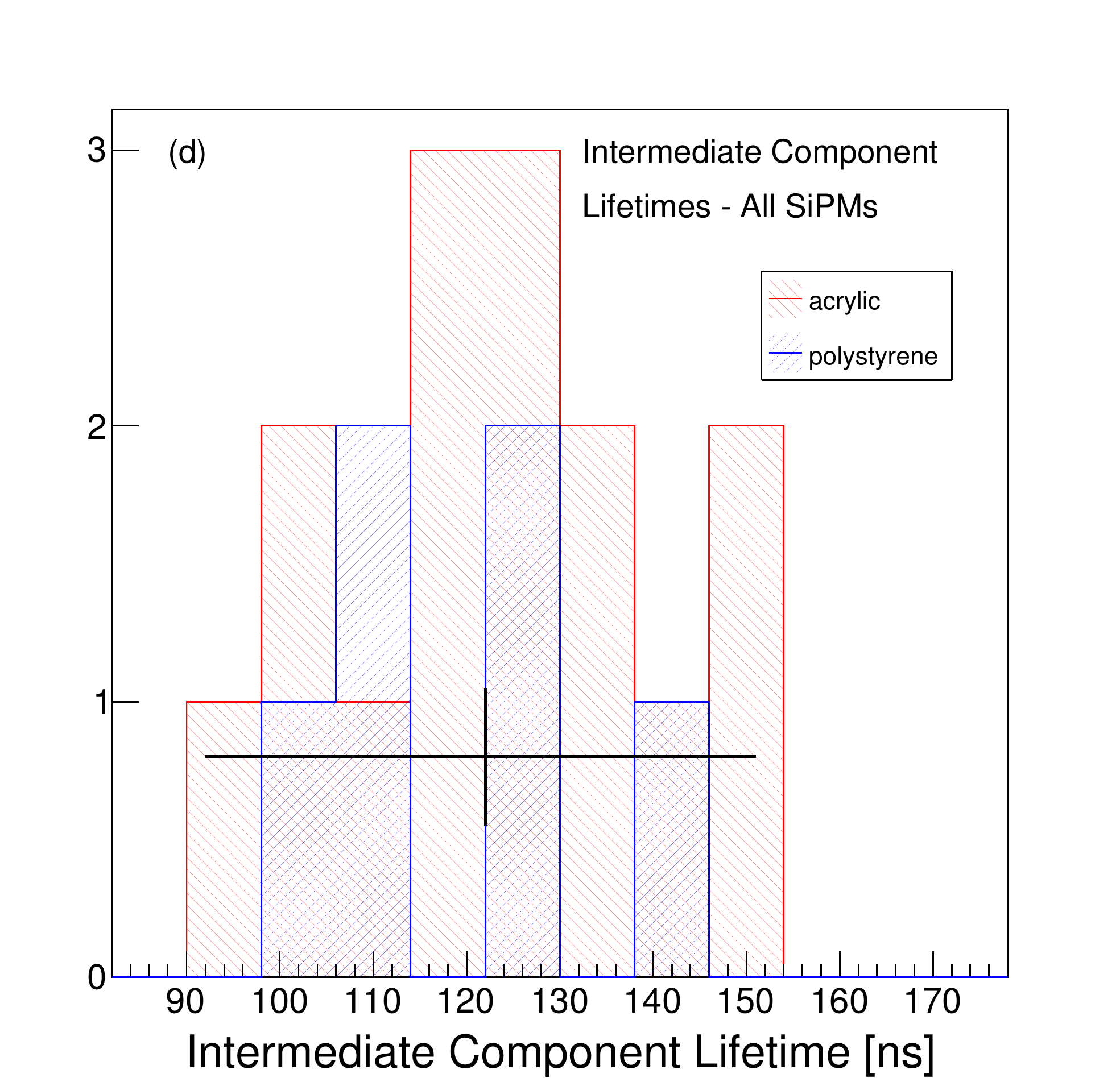}
    \caption{Histograms of the fit parameters $\tau_{\text{E}}$, $\tau_{\text{L}}$, $F_{\text{E}}$, and $\tau_{\text{I}}$ for the total set of SiPMs, separated by the composition of the light guide on which they are mounted.  (a) early component lifetime, $\tau_{\text{E}}$,  (b) late component lifetime, $\tau_{\text{L}}$, (c) percent of early light to total light, $F_{\text{E}}$,  (d) intermediate component lifetime, $\tau_{\text{I}}$.  Superposed on the histograms for $\tau_{\text{L}}$, $F_{\text{E}}$, and $\tau_{\text{I}}$ are the means for the parameters from Table~\protect\ref{tab:4comp}.  The horizontal line through the mean gives the range of the fit values from  Appendix~\protect\ref{sec:AppendixA}.}
    \label{fig:AllSiPMs}
  \end{center}
\end{figure}

Fig.~\ref{fig:AllSiPMs} shows histograms for the model fit parameters $\tau_{\text{E}}$, $\tau_{\text{L}}$, $F_{\text{E}}$, and $\tau_{\text{I}}$ for the total set of SiPMs.  (For two of the SiPM models on an acrylic light guide, it was necessary to fix the lifetime of the fourth component at the mean $\tau_4$ in Table~\ref{tab:4comp} to find an acceptable fit.) 
Two histograms are superposed for each fit parameter in Fig.~\ref{fig:AllSiPMs}, one for acrylic light guides (red) and one for polystyrene light guides (blue).  Also superposed on the histograms for $\tau_{\text{L}}$, $F_{\text{E}}$, and $\tau_{\text{I}}$ are the means for the parameter computed from Table~\ref{tab:4comp}.  The horizontal line through the mean gives the range of the fit values from the tables in Appendix~\ref{sec:AppendixA}.  

Fig.~\ref{fig:AllSiPMs}~(a) shows a clear separation of $\tau_{\text{E}}$ that depends on the composition of the light guide.  As in Table~\ref{tab:4comp}, SiPMs on polystyrene light guides exhibit longer lifetimes associated with the early-light component than SiPMs on acrylic.  This difference is likely related to an additional wavelength-shifting response associated with the doped polystyrene.  Figs.~\ref{fig:AllSiPMs}~(b) and (c) show that the parameters $\tau_{\text{L}}$ and $F_{\text{E}}$ exhibit no dependence on the composition of the light guides.  These are parameters measuring characteristics of liquid argon scintillation independent of the detectors.  The means for the high quality data in Table~\ref{tab:4comp} reasonably represent an average value for all the SiPMs.   Their range shows that higher quality data does improve the confidence in the mean values.  Fig.~\ref{fig:AllSiPMs}~(d) shows that the lifetime of the intermediate state, $\tau_{\text{I}}$, does not depend on the composition of the light guides but nonetheless has a fairly broad distribution.  Whether its origin is instrumental or associated with scintillation from LAr is not clear.  

\FloatBarrier
\section{Composite Model Description for the Illumination Functions}
\label{sec:PhysicalModel}

The phenomenological models presented in \S\ref{sec:4comp} find early and late components of LAr scintillation light that can be identified with the singlet and triplet decays from the $\text{Ar}_2^*$ dimer. However, these models leave unexplained the origins of the intermediate component seen in all the light guides and the long $\sim$6.6~$\mu$s component found in the acrylic light guides.  The phenomenological model, therefore, has been recast into a "composite model" which proposes that the scintillation photons from singlet and triplet decays of the $\text{Ar}_2^*$ dimer excite multiple decay modes in the WLS. It is then the convolution of the LAr emission function with the WLS response function that is used to fit the illumination function $I^k(t_i)$.

%The two known modes of scintillation light emission from LAr are the singlet and triplet $\text{Ar}_2^*$ states. Emission from TPB through a fast $\sim$1~ns component has been detected~\cite{bib:TPBdecay}, as have multiple delayed emission components whose decay constants can reach several microseconds~\cite{bib:Segreto}. 
%The phenomenological model of \S\ref{sec:4comp}, therefore, was recast into a form which better associates all observed components with physical processes. This ``physical model'' is again based on EMG functions Eq.~(\ref{eq:ExpModGauss}) since they have been shown to provide an excellent description of the shape of $I^k(t_i)$. Unlike the phenomenological model Eq.~(\ref{eq:response}), which sums a number of EMG components to find the best fit, the composite model proposes that the scintillation photons from singlet and triplet decays of the $\text{Ar}_2^*$ dimer excite multiple decay modes in the WLS. It is then the convolution of the LAr emission function with the WLS response function that is used to fit the illumination function $I^k(t_i)$.

\subsection{Composite Model Parameterization}
\label{sec:physicalModelParameterization}

The composite model is again based on EMG functions Eq.~(\ref{eq:ExpModGauss}) since they have been shown to provide an excellent description of the shape of the illumination functions $I^k(t_i)$.  The composite model for the illumination functions $I^k(t_i)$ is the convolution of two LAr scintillation modes and $n_{ws}$ WLS emission modes. In analogy with Eq.~(\ref{eq:response}),  
\begin{equation}
I^k(t_i) = \sum_{j=\text{S},\text{T}} A_j  \,  \sum_{l}^{n_{ws}} f_{l} \frac{1}{\tau_{l}-\tau_j} \left[ \tau_{l} \mathcal{E}(t_i,\tau_{l},w,t_m) - \tau_j \mathcal{E}(t_i,\tau_j,w,t_m) \right],
\label{eq:IlluminationModel}
\end{equation}
where the superscripts $k$ on the variables $A_j^k$, $f_l^k$, $\tau_j^k$, $\tau_l^k$, $w^k$, $t_m^k$ have been suppressed for clarity.  The $j$-sum includes the singlet and triplet state $\text{Ar}_2^*$ decays.  The $l$-sum runs over the number of WLS emission components, $n_{ws}$. The parameters $f_l$ are the fractions of visible emission contributed by each WLS emission mode to the illumination function. 
%In the case of one WLS component with $f_{\text{W1}} = 1$ and $\tau_{\text{W1}} = 0$~ns, Eq.~(\ref{eq:IlluminationModel}) reduces to Eq.~(\ref{eq:response}).

The composite model assumes there are three VUV wavelength shifting contributions ($n_{ws}=3$) in addition to the two $\text{Ar}_2^*$ decays. That makes 12 parameters in this model.  Emission from TPB through a fast $\sim$1~ns component has been detected~\cite{bib:TPBdecay}, as have delayed emission components whose decay constants can reach several microseconds~\cite{bib:Segreto}.  Since the timing resolution of the experiment is not capable of precisely measuring the fast component of WLS emission, the decay constant for the fastest WLS component $n_{ws} = 1$ is set to $\tau_{ws1} = 1$~ns for both TPB~\cite{bib:TPBdecay} and bis-MSB.  It is assumed that both organic scintillators have similar fast components. It is also assumed that all wavelength-shifted photons that contribute to $I^k(t_i)$ come from one of the WLS decay modes, which imposes the normalization condition $\sum_l f_l = 1$. With these assumptions, the number of fit parameters then reduces to 10.  Consequently, both the models of Eq.~(\ref{eq:response}) and Eq.~(\ref{eq:IlluminationModel}) have the same number of fit parameters.  

\subsection{Fit Results}
\label{sec:fitResultsPhysicalModel}

\begin{figure}[ht]
  \begin{center}
    \includegraphics[width=.65\columnwidth]{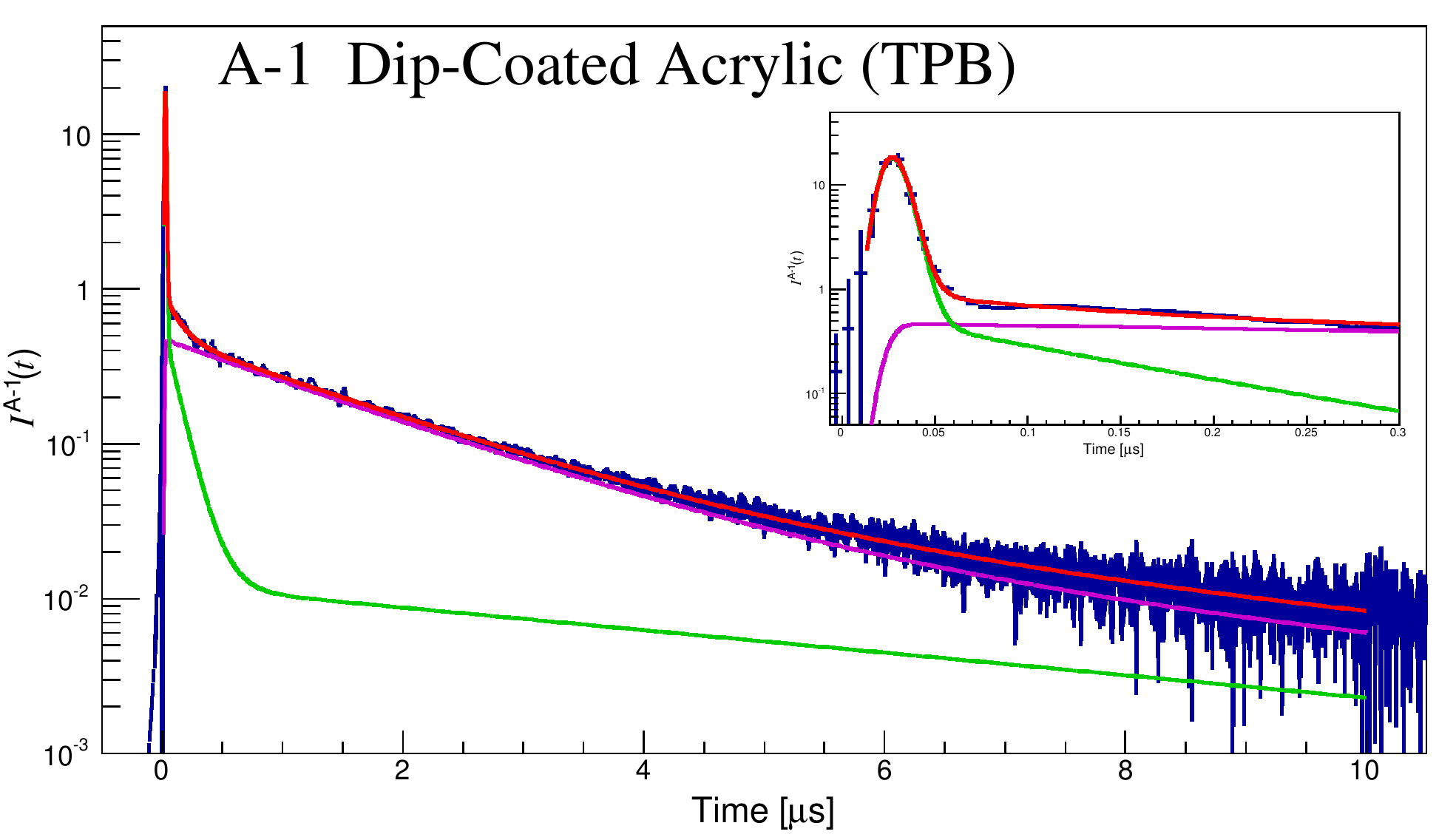}
    \caption{The illumination function $I^{A-1}$ observed by SiPM 1 on light guide A. The composite model is superposed in red.  Shown individually are the singlet component in green and the triplet component in magenta.}
    \label{fig:AltScintFit}
  \end{center}
\end{figure}

The fits to $I^k(t_i)$ with the composite model are found in Appendix~\ref{sec:AppendixB}. 
Their $\chi^2/N_{\text{DF}}$ values show that the composite model fits the illumination functions for the SiPMs as well as Eq.~(\ref{eq:response}).
As an example, Fig.~\ref{fig:AltScintFit} shows the composite model fit to the illumination function for SiPM A-1 superposed on its illumination function $I^{A-1}$.
The fit clearly describes the illumination function as well as the phenomenological model in Fig.~ \ref{fig:4comp-ScintFits}.  This is the expected result since the composite model recasts the components of the phenomenological model into components associated with physical processes.

Table~\ref{tab:AltPars1} gives the mean and standard deviation for the decay constants found for the three SiPMs on each light guide. The singlet and triplet decay constants, which characterize the decay of the $ \text{Ar}_2^*$ dimer, are consistent with those in Table~\ref{tab:4comp}. 
\begin{table}[ht]
  \begin{center}
    \caption{The mean and standard deviation for the decay constants on each light guide in the composite model.}
    \vspace{0.2em}
    \label{tab:AltPars1}
   \begin{tabular}{| c  |r@{\,$\pm$\,}l r@{\,$\pm$\,}l r@{\,$\pm$\,}l r@{\,$\pm$\,}l  |}
    \hline
    \hline
    Light Guide & \multicolumn{2}{c}{$\tau_{\text{S}}$ [ns]} & \multicolumn{2}{c}{$\tau_{\text{T}}$ [ns]} & \multicolumn{2}{c}{$\tau_{\text{w2}}$  [ns]} & \multicolumn{2}{c|}{$\tau_{\text{w3}}$  [ns]}  \\
    \hline
    A & ~ 4.4 & 2.0 & 1503 &  41    & 132 & 11 & 6658 & 1160   \\
    B & 4.8 & 1.1 & 1536 &  40 & 130 & 12 & 6608 & 1039  \\
    C & 4.7 & 1.0 & 1524 &  47 & 105 & 9 & 7777 & 1605  \\
    D & 8.0 & 1.1 & 1524 & 6 & 131 & 7 & \multicolumn{2}{c|}{---}  \\
   \hline
    \hline
    \end{tabular}
  \end{center}
\end{table}
Fitting this model to the 10 SiPMs excluded from the main analysis yields decay constants consistent with the SiPMs in Table~\ref{tab:AltPars1}. The composite model also confirms that the $\sim$6.6~$\mu$s component in light guide D makes a negligible contribution to the illumination function. 
These results suggest that the intermediate and long components in both models are due to a complicated delayed response to excitation by VUV photons of the TPB and bis-MSB embedded in a plastic matrix.

Table~\ref{tab:AltPars2} shows how the singlet light and the early light are related in the two models.
The second column of the table gives $f_{\text{w1}} = 1- (f_{\text{w2}}+f_{\text{w3}})$, the fraction of the scintillation light radiated by the fast emission component of the WLS, and the third column gives $F_{\text{S}} = A_{\text{S}}/(A_{\text{S}}+A_{\text{T}})$, the fraction of singlet light to the total (singlet + triplet) light from LAr scintillation.
\begin{table}[ht]
  \begin{center}
    \caption{The relation of the singlet emission to the early light emission in the phenomenolgical model.}
    \vspace{0.2em}
    \label{tab:AltPars2}
%    \begin{tabular}{| c r@{\,$\pm$\,}l r@{\,$\pm$\,}l r@{\,$\pm$\,}l r@{\,$\pm$\,}l | r@{\,$\pm$\,}l |}
    \begin{tabular}{| c | c@{\,$\pm$\,}l  r@{\,$\pm$\,}l |  r@{\,$\pm$\,}l |}
    \hline
    \hline
    Light Guide & \multicolumn{2}{c}{$f_{\text{w1}}$ } & \multicolumn{2}{c}{$F_{\text{S}}$ [\%]} & \multicolumn{2}{|c|}{$F_{\text{E}}~[\%] = f_{\text{w1}}  \times F_{\text{S}}  $} \\
    \hline
    A & ~~0.71 & 0.02 & ~~35.3 & 2.5 ~ & ~~~~~~25.1 & 1.9 \\
    B & ~~0.70 & 0.02 & 34.0 & 2.5 & ~~~~~~23.8 & 1.9 \\
    C & ~~0.69 & 0.02 & 39.5 & 3.0 & ~~~~~~27.3 & 2.2 \\
    D & ~~0.72 & 0.03 & 36.6 & 1.8 & ~~~~~~26.4 & 1.7 \\
    \hline
    \hline
    \end{tabular}
  \end{center}
\end{table}
The last column is the product of $f_{\text{w1}}$ and $F_{\text{S}}$, which equals the early component of the scintillation light $F_{\text{E}}$ in the phenomenological model of Table~\ref{tab:4comp}.  This table  
demonstrates that the early component of the scintillation light in the phenomenological model, which is often identified as the total emission from the singlet state decays, is actually only about 70\% of these decays.  The remaining 30\% is seen at later times as radiation from the two components with longer decay constants $\tau_{\text{w2}}$ and $\tau_{\text{w3}}$ of WLS emission.  
As in the phenomenological model, light guide D exhibits a fast component with a longer lifetime of 8~ns, which suggests an additional delayed response from the interaction of TPB with polystyrene. 
Such a response has been reported with a lifetime of 8-16~ns~\cite{bib:Swank}. 
While this substructure could not be distinguished with the time resolution of this experiment, a preliminary study that simply fixed $\tau_{\text{w3}} = 12$~ns resulted in a fit that recovered $\tau_{\text{S}} \approx 5$~ns. 
Without a more accurate description of the non-exponential response of TPB and bis-MSB, however, and their interaction with acrylic and polystyrene, these delayed components cannot be more precisely determined by this study. 

%Interestingly, by distributing the contributions from the $\sim$100~ns and $\sim$6.6~$\mu$s effects among the two LAr scintillation components in this model, the fractional contribution from the singlet component $F_{\text{S}} = A_{\text{S}}/(A_{\text{S}}+A_{\text{T}})$ becomes $\sim$37\%. This is because $\sim$30\% of the singlet signal incident on the detector is observed at later times due to delayed emission from the WLS and is instead attributed to the other components within the three- or four-component models. However, without a more precise description of the non-exponential response of TPB and bis-MSB and their interaction with acrylic and polystyrene the attribution of these delayed components cannot be more precicely determined by this study. The range of values for $F_{\text{S}}$ extracted from this model reflects this systematic uncertainty. For comparison with the phenomenological parameterization of \S\ref{sec:4comp} the final column calculates the fraction of signal arriving as the early component (singlet-state VUV photons detected via the fast WLS process). All values are consistent with those of Table~\ref{tab:Results}.

%\newpage

\FloatBarrier
\section{Discussion}
\label{sec:Discussion}

Table~\ref{tab:Results} summarizes several results from the fits to the illumination function.  The phenomenological parameter $\tau_{\text{L}}$ and the composite model parameter $\tau_{\text{T}}$ are the constants characterizing the decay of the triplet state of the $ \text{Ar}_2^*$ dimer.  This is the first direct measurement of $\tau_{\text{T}}$ for cosmic-ray muons.  The triplet state decay constant for LAr has been been found to fall in the range $\tau_{\text{T}} = 1.2 - 1.6~\mu$s for electrons~\cite{bib:Hitachi1,bib:pulseShape2,bib:N2Contamination,bib:O2Contamination,bib:pulseShape,bib:LArScint}.
\begin{table}[ht]
  \begin{center}
    \caption{Model Comparisons and the Computation of the Prompt Fraction of Scintillation Light}
    \vspace{0.2em}
    \label{tab:Results}
    \begin{tabular}{ | c | c | c | c | c | c | c |}
      \hline
      \hline
      Light &             & \multicolumn{2}{c|}{4-Component Model} & \multicolumn{2}{c|}{Composite Model} & \\ \cline{3-6}
      Guide & Technology & $\tau_{\text{L}}$ [ns] & $F_{\text{E}}$ [\%] & $\tau_{\text{T}}$ [ns] & 
      $F_{\text{S}}$ [\%]  & $F_{\text{prompt}}$   \\
      \hline
      A & acrylic, dipped     & 1517& 25.0 & 1503 & 35.3 & 0.31 \\
         &  TPB   &  & &  &  &  \\
      B & acrylic, dipped  & 1528 & 23.6 & 1536 & 34.0 & 0.29 \\
         & bis-MSB  &  &  &  & &  \\
      C & acrylic, doped       & 1527 & 27.1 & 1524 & 39.5 & 0.33 \\
        & bis-MSB       &  &  &  &  &  \\
      D & poly, doped       & 1524 & 25.8 & 1524 & 36.6 & 0.35 \\
         & TPB     &  &  &  &  &  \\
      \hline
      \multicolumn{2}{| r |}{mean $\pm$ st.dev.} & 1524 $\pm$ 5 & 25.4 $\pm$ 1.5 & 1522 $\pm$ 13 & 36.4 $\pm$ 2.4 
       & 0.32 $\pm$ 0.03\\
      \hline
     \hline
    \end{tabular}
  \end{center}
\end{table}
The value of $\tau_{\text{T}} = 1.52~\mu$s found in this investigation fits well within the range for electrons and is consistent with the picture of electrons and muons exciting LAr similarly.  However, the $e^{-}$ measurements found in the literature for the most part do not have overlapping error bars.  Possible explanations for this wide range include error limits that were underestimated in some studies and/or improper corrections for contaminants.  

This investigation shows that the identification of the ``early'' light component in the phenomenological model with the total light from singlet decays is an underestimate.  From the values of $F_{\text{E}}$ and $F_{\text{S}}$ in the table, approximately 30\% of the singlet light is emitted by the WLS through processes with long decay constants.  An indirect measurement of the fraction of early light for cosmic-ray muons in LAr has been reported by ICARUS~\cite{bib:ICARUS1}.  In that experiment, the fraction of early light was found to be 0.24~$\pm$~0.08, which matches well with $F_{\text{E}}$ in Table~\ref{tab:Results}.  There are several parameters in the literature other than $F_{\text{E}}$ that have been used to characterize the early light.  These can be computed using Tables~\ref{tab:4compBySiPM-A}--\ref{tab:3compBySiPM-D}. 

The parameter $F_{\text{E}}$, however, is often not what experiments measure.  Dark matter experiments, for example, use the parameter $F_{\text{prompt}}$, the fraction of light coming in a short time window after the trigger compared with the light in the total recorded waveform, to distinguish $\beta$'s from nuclear recoils and $\alpha$'s~\cite{bib:pulseShape2,bib:pulseShape10,bib:pulseShape,bib:WARP,bib:boulay}.  This parameter has also been used to separate muons from more heavily ionizing $\alpha$'s~\cite{bib:Hitachi1,bib:pulseShape,bib:carvalho}.  

The model fits to the illumination functions can be used to estimate $F_{\text{prompt}}$ for cosmic-ray muons by computing Eq.~(\ref{eq:deconvolution}) over the experiment's short time window in the numerator and again over the total waveform in the denominator.  Since the response function and time windows for every experiment are different, $F_{\text{prompt}}$ is clearly experiment dependent.  Here a simplified calculation for this experiment is presented in Table~\ref{tab:Results} in which the measurements are made with an ideal detector with $\delta$-function response, a short ``early'' time window of length $t_\ast$, and a long time window for the total waveform of length $t_f$.  Then the calculation of $F_{\text{prompt}}$ simplifies to 
\begin{equation}
F_{\text{prompt}} = \frac{\int_0^{t_\ast} I^k(t) dt }{\int_0^{t_f} I^k(t) dt} \approx
\sum_{i=1}^{n(t_\ast)} I^k(t_i) / \sum_{i=1}^{n(t_f)}  I^k(t_i),
\label{eq:Fprompt}
\end{equation}
where $n(t_\ast)$ is the number of time bins spanning $t_\ast$ and $n(t_f)$ is the number of time bins spanning $t_f$.  For this experiment, $t_\ast = 133$~ns as in Fig.~\ref{fig:NoiseFigs} and $n(t_\ast) = 20$; $t_f = 10~\mu$s and $n(t_f) = 1500$.  The results for $F_{\text{prompt}}$ in Table~\ref{tab:Results} are not strongly dependent on the light guide technologies used in this investigation.  

Several dark matter and double $\beta$-decay experiments have used $F_{\text{prompt}}$ to discriminate electron recoil backgrounds from nuclear/nucleus recoils.  The values of $t_\ast$ and $t_f$ for the calculation of $F_{\text{prompt}}$ were all optimized by the individual experiments.  Nevertheless, the values of $F_{\text{prompt}}$ found in these experiments are consistent with Table~\ref{tab:Results}, suggesting that the prompt fraction is not strongly dependent on the exact choices for $t_\ast$ and $t_f$.  This consistency also suggests that the scintillation emission profiles for electrons and muons in LAr are similar, as expected.  
Among the experimental values for electron recoil: $F_{\text{prompt}} \sim$0.3 (energy dependent, 0.39 at 5 keV and 0.28 at 30 keV) with $t_\ast =50$~ns and $t_f = 9~\mu$s~\cite{bib:pulseShape2}; $F_{\text{prompt}} \sim$0.3 with $t_\ast =90$~ns, $t_f = 7~\mu$s~\cite{bib:pulseShape10}; $F_{\text{prompt}} \sim 0.28-0.29$ for 2 MeV betas/$\gamma$s with $t_\ast =40$~ns, $t_f = 6~\mu$s~\cite{bib:pulseShape}; $F_{\text{prompt}} = 0.3$ with $t_\ast =120$~ns and $t_f = 7~\mu$s~\cite{bib:WARP}; $F_{\text{prompt}} = 0.3$ with $t_\ast =100$~ns and $t_f = 9~\mu$s~\cite{bib:boulay}.

Experiments that fit two-component models to the time structure of the LAr scintillation signals from electrons and muons report values in the range 0.25 -- 0.30 for the ratio of ``early/late light''~\cite{bib:Hitachi1,bib:pulseShape2,bib:pulseShape}.  However, two-component models have been shown to be inadequate in describing the illumination functions~(\S\ref{sec:4comp}).  Further, it is not clear how to compute this fraction from the four-component models presented here.   Summing the amplitudes of the long components and comparing the sum to $A_{\text{E}}$ for the phenomenological models gives results in the range 0.31 -- 0.39.  

%~~~~~~~~~~~~~~~~~~~~~~~~~~~~~~
\section{Summary}
\label{sec:Summary}

This paper reports the results of an experiment to directly measure the time-resolved scintillation signal from the passage of cosmic-ray muons through liquid argon.  Scintillation light from these muons is of value to studies of weakly-interacting particles in neutrino experiments and dark matter searches. The experiment took place at the large liquid argon test facility TallBo at Fermilab.  The experimental apparatus consisted of DUNE prototype photon detector paddles that are made up of four light guides that capture VUV scintillation photons, convert the photons into the optical, and then channel them to the SiPM photosensors at one end.  The waveforms acquired by the SiPMs over 10~$\mu$s were read out by prototype DUNE electronics.  There were several technologies tested in the PD paddles.  The light guides were made of either UVT acrylic or polystyrene.  The waveshifter incorporated into the light guides to convert the VUV scintillation photons was either TPB or bis-MSB.  The photodetectors were SensL SiPMs. 

The deconvolution of the detector response function from the waveforms generated by cosmic-ray muons yields the illumination function, or the time sequence of scintillation photons incident on the light guides.
Two models were presented for the illumination functions, a four-parameter phenomenological model and a four-component physically-motivated model.  Both models find that the decay of the triplet state of the $\text{Ar}_{\text{2}}^*$ dimer to be $ \tau_{\text{T}}  = 1.52$~$\mu$s.  This is the first direct measurement of $\tau_{\text{T}}$ for cosmic-ray muons and it falls well within the range of measurements for electrons, making it consistent with the picture of electrons and muons exciting LAr similarly.  
In addition, the identification of the early light fraction in the phenomenological model, $F_{\text{E}}\approx$~25\% of the signal, with the total light from singlet decays is an underestimate.  
The total fraction of singlet light is $F_{\text{S}} \approx$ 36\%, where the increase over $F_{\text{E}}$ is emitted by the WLS through processes with long decay constants.

The parameter $F_{\text{prompt}}$ is used by dark matter and double $\beta$-decay experiments to discriminate electron recoil backgrounds from nuclear/nucleus recoils.  Calculations using both illumination function models reproduce the typical experimental values $\sim$0.3 quite well, again suggesting that the scintillation emission profiles for electrons and muons in LAr are similar.  Since the values of $F_{\text{prompt}}$ measured by these experiments were obtained with many different detector optimizations, this parameter seems to provide a robust metric for discriminating electrons and muons from more heavily ionizing particles.

\section{Acknowledgments}

This work was partially supported by the Office of High Energy Physics of the DOE with grant 
DE-SC0010120 to Indiana University and by Brookhaven National Laboratory
with grant 240296-A3 to Indiana University. The authors wish to thank the many people who helped make this work possible. At IU: B.~Adams, B.~Baptista, B.~Baugh,  M.~Gebhard, M.~Lang, J.~Musser, P.~Smith, J.~Urheim.  At MIT:  L.~Bugel, J.~Conrad, B.~Jones, Z.~Moss, M.~Toups, T.~Wongjirad.  At Fermilab: R.~Davis, K.~Hardin, M.~Johnson, B.~Miner, S.~Pordes, B.~Rebel, M.~Ruschman.  At ANL: J.~Anderson, P.~DeLurgio, G.~Drake, V.~Guarino, A.~Kreps, M.~Oberling.  At Colorado State: D.~Adams, N.~Buchanan, F.~Craft, T.~Cummings, J.~Jablonski, D.~Warner, R.~Wasserman.

\bibliography{BarResponseJINST}
\bibliographystyle{JHEP}

\clearpage
\appendix
\section*{Appendices}
\addcontentsline{toc}{section}{Appendices}
\renewcommand{\thesubsection}{\Alph{subsection}}
%\FloatBarrier
\subsection{Phenomenological Model Fit Results}
\label{sec:AppendixA}

\begin{table}[ht]
  \begin{center}        
    \caption{Four-component fit results, light guide A.}
    \vspace{0.2em}
    \label{tab:4compBySiPM-A}
    \begin{tabular}{l r@{\,$\pm$\,}l r@{\,$\pm$\,}l r@{\,$\pm$\,}l }
      \hline
      & \multicolumn{2}{c}{SiPM A-0} & \multicolumn{2}{c}{SiPM A-1} & \multicolumn{2}{c}{SiPM A-2} \\
      \hline
      \hline
      $\tau_{\text{E}}$ [ns] &  3.4 & 1.7 &  4.6 & 0.6 &  5.3 & 0.5 \\
      $\tau_{\text{I}}$ [ns] &  118 &   5 &  126 &   5 &  151 &   8 \\
      $\tau_{\text{L}}$ [ns] & 1523 &  21 & 1498 &  24 & 1488 &  25 \\
      $\tau_4$ [ns] & 7498 & 826 & 5963 & 539 & 6515 & 648 \\
      \hline
      $A_{\text{E}}$ & 0.280 & 0.022 & 0.331 & 0.029 & 0.285 & 0.019 \\
      $A_{\text{I}}$ & 0.044 & 0.001 & 0.053 & 0.002 & 0.049 & 0.002 \\
      $A_{\text{L}}$ & 0.601 & 0.011 & 0.671 & 0.016 & 0.592 & 0.012 \\
      $A_4$ & 0.209 & 0.006 & 0.246 & 0.012 & 0.215 & 0.009 \\
      \hline
      $w$ [ns]  &  6.5 & 1.0 &  6.2 & 1.2 &  5.8 & 0.8 \\
      $t_m$ [ns] & 24.4 & 1.5 & 23.3 & 1.0 & 22.7 & 0.7 \\
      \hline
      $\chi^2/N_{\text{DF}}$ & \multicolumn{2}{c}{0.57} & \multicolumn{2}{c}{0.59} & \multicolumn{2}{c}{0.67} \\
      \hline
      \hline
    \end{tabular}
  \end{center}
\end{table}

\begin{table}[ht]
  \begin{center}
    \caption{Four-component fit results, light guide B.}
    \vspace{0.2em}
    \label{tab:4compBySiPM-B}
    \begin{tabular}{l r@{\,$\pm$\,}l r@{\,$\pm$\,}l r@{\,$\pm$\,}l }
      \hline
      & \multicolumn{2}{c}{SiPM B-0} & \multicolumn{2}{c}{SiPM B-1} & \multicolumn{2}{c}{SiPM B-2} \\
      \hline
      \hline
      $\tau_{\text{E}}$ [ns] &  4.8 & 0.7 &  4.9 & 0.5 &  5.1 & 0.5 \\
      $\tau_{\text{I}}$ [ns] &  125 &   8 &  128 &   6 &  138 &   7 \\
      $\tau_{\text{L}}$ [ns] & 1518 &  23 & 1566 &  22 & 1524 &  24 \\
      $\tau_4$ [ns] & 6409 & 453 & 7208 & 449 & 6209 & 420 \\
      \hline
      $A_{\text{E}}$ & 0.514 & 0.050 & 0.546 & 0.039 & 0.503 & 0.033 \\
      $A_{\text{I}}$ & 0.089 & 0.004 & 0.094 & 0.003 & 0.077 & 0.003 \\
      $A_{\text{L}}$ & 1.142 & 0.022 & 1.245 & 0.024 & 1.080 & 0.024 \\
      $A_4$ & 0.453 & 0.016 & 0.449 & 0.014 & 0.430 & 0.018 \\
      \hline
      $w$ [ns]  &  5.5 & 1.0 &  5.9 & 0.8 &  6.0 & 0.8 \\
      $t_m$ [ns] & 22.2 & 1.0 & 22.3 & 0.8 & 22.8 & 0.7 \\
      \hline
      $\chi^2/N_{\text{DF}}$ & \multicolumn{2}{c}{0.46} & \multicolumn{2}{c}{0.47} & \multicolumn{2}{c}{0.54} \\
      \hline
      \hline
    \end{tabular}
  \end{center}
\end{table}

\begin{table}[ht]
  \begin{center}
    \caption{Four-component fit results, light guide C.}
    \vspace{0.2em}
    \label{tab:4compBySiPM-C}
    \begin{tabular}{l r@{\,$\pm$\,}l r@{\,$\pm$\,}l r@{\,$\pm$\,}l }
      \hline
      & \multicolumn{2}{c}{SiPM C-0} & \multicolumn{2}{c}{SiPM C-1} & \multicolumn{2}{c}{SiPM C-2} \\
      \hline
      \hline
      $\tau_{\text{E}}$ [ns] &  5.8 & 0.5 &  4.6 &  0.6 &  4.1 &  3.1 \\
      $\tau_{\text{I}}$ [ns] &  104 &   5 &   92 &    3 &  120 &    7 \\
      $\tau_{\text{L}}$ [ns] & 1515 &  31 & 1507 &   22 & 1551 &   27 \\
      $\tau_4$ [ns] & 6160 & 594 & 8441 &  938 & 8732 & 1135 \\
      \hline
      $A_{\text{E}}$ & 0.158 & 0.011 & 0.228 & 0.017 & 0.235 & 0.027 \\
      $A_{\text{I}}$ & 0.029 & 0.001 & 0.039 & 0.001 & 0.041 & 0.002 \\
      $A_{\text{L}}$ & 0.261 & 0.008 & 0.414 & 0.008 & 0.408 & 0.008 \\
      $A_4$ & 0.116 & 0.006 & 0.183 & 0.003 & 0.172 & 0.003 \\
      \hline
      $w$ [ns]  &  5.4 & 0.7 &  5.8 & 0.8 &  5.6 & 1.5 \\
      $t_m$ [ns] & 21.7 & 0.7 & 22.1 & 0.8 & 22.4 & 2.3 \\
      \hline
      $\chi^2/N_{\text{DF}}$ & \multicolumn{2}{c}{0.84} & \multicolumn{2}{c}{0.72} & \multicolumn{2}{c}{0.88} \\
      \hline
      \hline
    \end{tabular}
  \end{center}
\end{table}

\begin{table}[ht]
  \begin{center}
    \caption{Three-component fit results, light guide D.}
    \vspace{0.2em}
    \label{tab:3compBySiPM-D}
    \begin{tabular}{l r@{\,$\pm$\,}l r@{\,$\pm$\,}l r@{\,$\pm$\,}l }
      \hline
      & \multicolumn{2}{c}{SiPM D-0} & \multicolumn{2}{c}{SiPM D-1} & \multicolumn{2}{c}{SiPM D-2} \\
      \hline
      \hline
      $\tau_{\text{E}}$ [ns] &  7.9 & 0.4 &  8.2 &  0.6 &  8.2 &  0.9 \\
      $\tau_{\text{I}}$ [ns] &  139 &   4 &  127 &    4 &  127 &    4 \\
      $\tau_{\text{L}}$ [ns] & 1523 &   4 & 1508 &    3 & 1541 &    4 \\
      \hline
      $A_{\text{E}}$ & 0.418 & 0.020 & 0.472 & 0.033 & 0.403 & 0.034 \\
      $A_{\text{I}}$ & 0.153 & 0.003 & 0.162 & 0.003 & 0.148 & 0.003 \\
      $A_{\text{L}}$ & 1.082 & 0.002 & 1.198 & 0.002 & 0.957 & 0.002 \\
      \hline
      $w$ [ns]  &  5.3 & 0.8 &  5.5 & 1.1 &  5.6 & 1.2 \\
      $t_m$ [ns] & 23.1 & 0.6 & 23.1 & 0.8 & 22.5 & 1.0 \\
      \hline
      $\chi^2/N_{\text{DF}}$ & \multicolumn{2}{c}{0.81} & \multicolumn{2}{c}{0.58} & \multicolumn{2}{c}{0.73} \\
      \hline
      \hline
    \end{tabular}
  \end{center}
\end{table}

\clearpage
%\FloatBarrier
\subsection{Composite Model Fit Results}
\label{sec:AppendixB}

\begin{table}[ht]
  \begin{center}        
    \caption{Composite model fit results, light guide A.}
    \vspace{0.2em}
    \label{tab:AltFitsBySiPM-A}
    \begin{tabular}{l r@{\,$\pm$\,}l r@{\,$\pm$\,}l r@{\,$\pm$\,}l }
      \hline
      & \multicolumn{2}{c}{SiPM A-0} & \multicolumn{2}{c}{SiPM A-1} & \multicolumn{2}{c}{SiPM A-2} \\
      \hline
      \hline
      $\tau_{\text{S}}$ [ns] &  3.3 & 1.8 &  4.6 & 0.7 &  5.2 & 0.5 \\
      $\tau_{\text{T}}$ [ns] & 1523 &  21 & 1498 &  24 & 1488 &  25 \\
      \hline
      $A_{\text{S}}$ & 0.395 & 0.025 & 0.466 & 0.031 & 0.410 & 0.020 \\
      $A_{\text{T}}$ & 0.740 & 0.007 & 0.835 & 0.007 & 0.730 & 0.006 \\
      \hline
 %     $\tau_{\text{w1}}$ [ns] &  \multicolumn{2}{c}{1}     &  \multicolumn{2}{c}{1}  &  \multicolumn{2}{c}{1}    \\
     $\tau_{\text{w2}}$ [ns] &  118 &   5 &  126 &   5 &  151 &   8 \\
      $\tau_{\text{w3}}$ [ns] & 7498 & 808 & 5963 & 519 & 6512 & 650 \\
      \hline
%      $f_{\text{w1}}$ &   &  &  &  &   &   \\
     $f_{\text{w2}}$ & 0.158 & 0.004 & 0.156 & 0.006 & 0.159 & 0.005 \\
      $f_{\text{w3}}$ & 0.129 & 0.010 & 0.133 & 0.011 & 0.145 & 0.010 \\
      \hline
      $w$ [ns]  &  6.3 & 1.0 &  5.9 & 1.1 &  5.5 & 0.8 \\
      $t_m$ [ns] & 22.6 & 1.5 & 21.5 & 0.9 & 20.9 & 0.7 \\
      \hline
      $\chi^2/N_{\text{DF}}$ & \multicolumn{2}{c}{0.57} & \multicolumn{2}{c}{0.59} & \multicolumn{2}{c}{0.67} \\
      \hline
      \hline
    \end{tabular}
  \end{center}
\end{table}

\begin{table}[ht]
  \begin{center}        
    \caption{Composite model fit results, light guide B.}
    \vspace{0.2em}
    \label{tab:AltFitsBySiPM-B}
    \begin{tabular}{l r@{\,$\pm$\,}l r@{\,$\pm$\,}l r@{\,$\pm$\,}l }
      \hline
      & \multicolumn{2}{c}{SiPM B-0} & \multicolumn{2}{c}{SiPM B-1} & \multicolumn{2}{c}{SiPM B-2} \\
      \hline
      \hline
      $\tau_{\text{S}}$ [ns] &  4.7 & 0.7 &  4.8 & 0.7 &  5.0 & 0.5 \\
      $\tau_{\text{T}}$ [ns] & 1518 &  23 & 1566 &  22 & 1524 &  24 \\
      \hline
      $A_{\text{S}}$ & 0.747 & 0.058 & 0.785 & 0.045 & 0.717 & 0.036 \\
      $A_{\text{T}}$ & 1.449 & 0.012 & 1.547 & 0.014 & 1.370 & 0.011 \\
      \hline
%     $\tau_{\text{w1}}$ [ns] &  \multicolumn{2}{c}{1}     &  \multicolumn{2}{c}{1}  &  \multicolumn{2}{c}{1}    \\
      $\tau_{\text{w2}}$ [ns] &  125 &   8 &  128 &   6 &  138 &   7 \\
      $\tau_{\text{w3}}$ [ns] & 6408 & 520 & 7207 & 731 & 6208 & 525 \\
      \hline
 %     $f_{\text{w1}}$ &   &  &  &  &   &   \\
     $f_{\text{w2}}$ & 0.171 & 0.005 & 0.162 & 0.004 & 0.170 & 0.006 \\
      $f_{\text{w3}}$ & 0.139 & 0.014 & 0.140 & 0.010 & 0.128 & 0.009 \\
      \hline
      $w$ [ns]  &  5.2 & 1.0 &  5.6 & 0.9 &  5.7 & 0.8 \\
      $t_m$ [ns] & 20.3 & 1.0 & 20.4 & 0.8 & 21.0 & 0.7 \\
      \hline
      $\chi^2/N_{\text{DF}}$ & \multicolumn{2}{c}{0.46} & \multicolumn{2}{c}{0.47} & \multicolumn{2}{c}{0.54} \\
      \hline
      \hline
    \end{tabular}
  \end{center}
\end{table}

\begin{table}[ht]
  \begin{center}        
    \caption{Composite model fit results, light guide C.}
    \vspace{0.2em}
    \label{tab:AltFitsBySiPM-C}
    \begin{tabular}{l r@{\,$\pm$\,}l r@{\,$\pm$\,}l r@{\,$\pm$\,}l }
      \hline
      & \multicolumn{2}{c}{SiPM C-0} & \multicolumn{2}{c}{SiPM C-1} & \multicolumn{2}{c}{SiPM C-2} \\
      \hline
      \hline
      $\tau_{\text{S}}$ [ns] &  5.7 & 0.5 &  4.5 & 0.6 &  4.0 & 0.7 \\
      $\tau_{\text{T}}$ [ns] & 1515 &  31 & 1507 &  21 & 1551 &  28 \\
      \hline
      $A_{\text{S}}$ & 0.230 & 0.012 & 0.333 & 0.019 & 0.344 & 0.032 \\
      $A_{\text{T}}$ & 0.334 & 0.004 & 0.531 & 0.007 & 0.513 & 0.008 \\
      \hline
 %     $\tau_{\text{w1}}$ [ns] &  \multicolumn{2}{c}{1}     &  \multicolumn{2}{c}{1}  &  \multicolumn{2}{c}{1}    \\
      $\tau_{\text{w2}}$ [ns] &  104 &   5 &   92 &   3 &  120 &    7 \\
      $\tau_{\text{w3}}$ [ns] & 6158 & 595 & 8440 & 918 & 8732 & 1174 \\
      \hline
 %     $f_{\text{w1}}$ &   &  &  &  &   &   \\
     $f_{\text{w2}}$  & 0.172 & 0.007 & 0.187 & 0.004 & 0.177 & 0.006 \\
      $f_{\text{w3}}$ & 0.136 & 0.009 & 0.128 & 0.008 & 0.135 & 0.016 \\
      \hline
      $w$ [ns]  &  5.1 & 0.8 &  5.5 & 0.8 &  5.4 & 1.7 \\
      $t_m$ [ns] & 19.8 & 0.8 & 20.2 & 0.8 & 20.6 & 2.7 \\
      \hline
      $\chi^2/N_{\text{DF}}$ & \multicolumn{2}{c}{0.84} & \multicolumn{2}{c}{0.72} & \multicolumn{2}{c}{0.88} \\
      \hline
      \hline
    \end{tabular}
  \end{center}
\end{table}

\begin{table}[ht]
  \begin{center}        
    \caption{Composite model fit results, light guide D.}
    \vspace{0.2em}
    \label{tab:AltFitsBySiPM-D}
    \begin{tabular}{l r@{\,$\pm$\,}l r@{\,$\pm$\,}l r@{\,$\pm$\,}l }
      \hline
      & \multicolumn{2}{c}{SiPM D-0} & \multicolumn{2}{c}{SiPM D-1} & \multicolumn{2}{c}{SiPM D-2} \\
      \hline
      \hline
      $\tau_{\text{S}}$ [ns] &  7.8 & 0.4 &  8.1 & 0.6 &  8.2 & 0.8 \\
      $\tau_{\text{T}}$ [ns] & 1523 &   4 & 1508 &   3 & 1541 &   4 \\
      \hline
      $A_{\text{S}}$ & 0.600 & 0.019 & 0.661 & 0.032 & 0.574 & 0.032 \\
      $A_{\text{T}}$ & 1.051 & 0.003 & 1.169 & 0.003 & 0.934 & 0.003 \\
      \hline
%      $\tau_{\text{w1}}$ [ns] &  \multicolumn{2}{c}{1}     &  \multicolumn{2}{c}{1}  &  \multicolumn{2}{c}{1}    \\
     $\tau_{\text{w2}}$ [ns] &  139 &   4 &  127 &   4 &  127 &    4 \\
      \hline
%     $f_{\text{w1}}$ &   &  &  &  &   &   \\
      $f_{\text{w2}}$ [ns] & 0.293 & 0.011 & 0.275 & 0.015 & 0.283 & 0.018 \\
      \hline
      $w$ [ns]  & 5.0 & 0.8 &  5.5 & 0.8 &  5.3 & 1.2 \\
      $t_m$ [ns] & 21.2 & 0.6 & 20.2 & 0.8 & 20.6 & 1.1 \\
      \hline
      $\chi^2/N_{\text{DF}}$ & \multicolumn{2}{c}{0.82} & \multicolumn{2}{c}{0.58} & \multicolumn{2}{c}{0.73} \\
      \hline
      \hline
    \end{tabular}
  \end{center}
\end{table}

\end{document}